\documentclass[%
 reprint,
amsmath,amssymb,
aps,
pre,
longbibliography
]{revtex4-1}

\usepackage{graphicx}
\usepackage{bm}
\usepackage{hyperref}

\usepackage{color}
\usepackage{graphicx}
\usepackage{subfigure}

\begin{document}

\title{
Lateral predictive coding revisited: Internal model, symmetry breaking, and response time
}

\author{Zhen-Ye Huang$^{1,2}$}
\author{Xin-Yi Fan$^{1,2}$}
\author{Jianwen Zhou$^{1,2}$}
\author{Hai-Jun Zhou$^{1,2,3}$}
\email{zhouhj@itp.ac.cn}
\affiliation{
$^1$CAS Key Laboratory for Theoretical Physics, Institute of Theoretical Physics, Chinese Academy of Sciences, Beijing 100190, China \\
  $^2$School of Physical Sciences, University of Chinese Academy of Sciences, Beijing 100049, China \\
$^3$MinJiang Collaborative Center for Theoretical Physics, MinJiang University, Fuzhou 350108, China
}

\date{\today}

\begin{abstract}
  Predictive coding is a promising theoretical framework in neuroscience for understanding information transmission and perception. It posits that the brain perceives the external world through internal models and updates these models under the guidance of prediction errors. Previous studies on predictive coding emphasized top-down feedback interactions in hierarchical multi-layered networks but largely ignored lateral recurrent interactions. We perform analytical and numerical investigations in this work on the effects of single-layer lateral interactions. We consider a simple predictive response dynamics and run it on the MNIST dataset of hand-written digits. We find that learning will generally break the interaction symmetry between peer neurons, and that high input correlation between two neurons does not necessarily bring strong direct interactions between them. The optimized network responds to familiar input signals much faster than to novel or random inputs, and it significantly reduces the correlations between the output states of pairs of neurons.
\end{abstract}


\maketitle


\section{Introduction}
\label{sec:intro}

The brain receives external and visceral sensory signals, extract useful information out of them, and make conscious or unconscious decisions on appropriate responses and actions all the time. Signal perception, decision making, and action triggering in the brain are carried out by a huge and complex network of interacting neurons, each of which integrates input signals and sends action spikes to other neurons. The brain with a size of about $2\%$ body mass consumes about $20\%$ of the body's metabolic energy, and it must be under strong evolutionary pressure to reduce energy consumption~\cite{Barlow-1961}. Predictive coding has been proposed as a general strategy to reduce the energy cost of information processing~\cite{Srinivasan-etal-1982,Huang-Rao-2011,Spratling-2017}.

Input signals to a neural network and the internal states of the network are produced by real-world physical or chemical processes, they are far from being completely random but are rich in internal structures at many temporal and spatial scales, and there is huge amount of regularity in their temporal and spatial structures. Regularity means redundancy and it could be exploited to facilitate information processing and to reduce energy cost. First, the signals received by spatially adjacent neurons at a given time are positively correlated, and this local spatial correlation could be exploited to reduce the magnitude of response of the receiving neurons~\cite{Srinivasan-etal-1982}. Second, the sequence of signals received at a given neuron are locally correlated in time, and this local temporal correlation could again be exploited to make prediction about future events~\cite{Montague-Sejnowski-1994,Palmer-etal-2015,Luczak-etal-2022}. Third, similar signal patterns (e.g., faces or objects) are repeatedly received by a neural network over longer time scales, and the stable hierarchical relationship among them could be exploited to build a hierarchical internal model of the world~\cite{Rao-Ballard-1999,Friston-2010,Keller-etal-2020,Aguilera-etal-2022}.

Predictive coding has came to be an influential and promising framework in theoretical neuroscience for understanding information transmission and perception. It posits that the brain builds an internal model to perceive the external world (and also the visceral world), and constantly transmits prediction error messages among its constituent neurons to guide the refinement of this model. Previous theoretical studies on predictive coding have paid great attention on top-down feedback mechanisms. The system was often modelled by a hierarchical network consisting of many layers of neurons. Special neurons were introduced into the hierarchical network to compute and transmit prediction errors between adjacent layers in the network, and Bayesian inference was employed to refine a hierarchical internal model~\cite{Huang-Rao-2011,Friston-2010,Aguilera-etal-2022,Jirsa-Sheheitli-2022}. For computational convenience,  the lateral recurrent interactions between the neurons located in the same layer of the network were usually ignored in these earlier models.  However, lateral interactions are ubiquitous in the biological brain. The mutual influences among the neurons in a single layer of the network strongly affect the state dynamics of these neurons in the short time scale, and they may then greatly affect perception and inference in the multi-layer network. Recent experimental and computational studies have demonstrated that the inclusion of within-layer interactions could dramatically change the performance of hierarchical neural networks (see, e.g., Refs.~\cite{Tang-etal-2018,Pang-etal-2021,Millidge-etal-2022}). Whether special prediction-error computing neurons really exist in the brain is also a widely debated issue~\cite{Mikulasch-etal-2022}.

The present work revisit the original concept of predictive coding within a single-layered neural network~\cite{Srinivasan-etal-1982,Pineda-1987,Foldiak-1990,Harpur-Prager-1996}. We consider a first-order differential equation of neurons $i$ responding to an external input with the help of peer neurons [Eq.~(\ref{eq:xevol}) and Fig.~\ref{fig:model}]. There is no need to introduce additional specialized neurons for computing prediction errors in our model. The internal state $x^i$ of neuron $i$ serves the dual role of a prediction error, while the combined effect $f_i(\bm{x})$ of other neurons to this neuron is interpreted as a prediction [Eq.~(\ref{eq:linmod})]. The synaptic weights $w_{i j}$ of lateral interactions from neuron $j$ to neuron $i$ are gradually optimized (on time scales much longer than that of the elementary response dynamics) to reduce the average squared prediction error. We implement a gradient descent algorithm to accomplish the task of synaptic weight adaption. Our theoretical derivation indicates that, as some of the synaptic weights deviate from being zero with learning, the symmetry of the synaptic weights is gradually lost ($w_{i j} \neq w_{j i}$).

We apply this predictive coding model to the MNIST dataset of hand-written digits. Our numerical results confirm the spontaneous breaking of synaptic weight symmetry, and they also demonstrate that high input correlation between two neurons does not necessarily mean there will be strong direct interactions between them in the optimized network. Other properties of the lateral predictive coding are also demonstrated, including the reduction of correlation among the responses of different neurons, and the attention mechanism of highlighting novel spots in the input signals. Of especial interest is that the response speed of the optimized perception system to familiar input signals is much faster than to unfamiliar or random inputs. We believe that lateral recurrent  interactions are indispensable in understanding predictive coding in biological nervous systems.

This paper is organized as follows. Section~\ref{sec:model} describes the fast-time-scale response dynamics and introduces the synaptic matrix of lateral interactions. Section~\ref{sec:adapt} defines the cost function to be minimized and derives the gradient descent iteration equations of the slow-time-scale adaptation of the synaptic weights. Section~\ref{sec:mnist} reports the main numerical results obtained on the hand-written digits dataset. Finally we conclude our work in section~\ref{sec:conclude}.

\section{Model}
\label{sec:model}

\subsection{Response dynamics and internal state}

\begin{figure}
\centering
\includegraphics[width=0.7\linewidth]{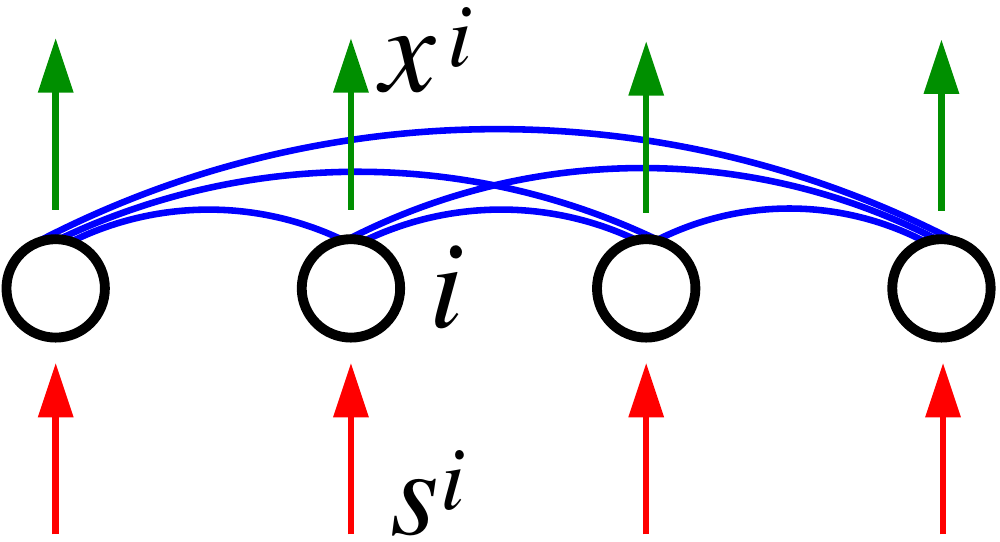}
\caption{
Lateral feedback interactions in a single layer of $N$ neurons. The input signal $s^i$ to a neuron $i$ is converted to an output signal $x^i$ by a quick response dynamics (\ref{eq:xevol}). Lateral interactions are indicated by the horizontal arcs. The interaction strengths of two neurons $i$ and $j$ are quantified by a pair of synaptic weights $w_{i j}$ and $w_{j i}$.
}
\label{fig:model}
\end{figure}

We focus attention on a single layer of neurons (Fig.~\ref{fig:model}). The $N$ neurons in this layer are affected by external inputs, and they are also mutually affected by lateral interactions~\cite{Srinivasan-etal-1982,Pineda-1987,Foldiak-1990,Harpur-Prager-1996}. We denote by $s^i$ the external input to a neuron $i$, and by $x^i$ the internal state of this neuron. If there is no external perturbation, the neurons stay in the quiescent state ($x^i=0$). Upon receiving an input signal $\bm{s} = (s^1, \ldots, s^N)^\top$, the internal state vector $\bm{x}=(x^1, \ldots, x^N)^\top$ is driven away from quiescence and reaches a new steady state quickly, on a time scale of milliseconds. We assume the following simple response dynamics,
\begin{equation}
\label{eq:xevol}
  \frac{ {\rm d} \bm{x}}{{\rm d} t} =
  \bm{s} - \bm{x} - \bm{f}(\bm{x} ) \; .
\end{equation}
%

The term $-\bm{x}$ on the right-hand side of this expression is the spontaneous relaxation, whose time constant is defined as unity. The vector function $\bm{f}(\bm{x}) = \bigl(f_1(\bm{x}), \ldots, f_N(\bm{x}) \bigr)^\top$ contains all the lateral feedback interactions between the $N$ neurons. These lateral interactions are generally nonlinear in a biological neural network. Here, for simplicity, we restrict the discussion to linear interactions and assume that the interaction function $f_i(\bm{x})$ has the following form
\begin{equation}
  f_i(\bm{x}) = 
  \sum\limits_{j\neq i} w_{i j} x^j \; ,
  \label{eq:linmod}
\end{equation}
where $w_{i j}$ is the synaptic weight of the directed interaction from neuron $j$ to neuron $i$. The synaptic weights define a lateral interaction matrix $\bm{W}$ as
\begin{equation}
\bm{W} = \left(
     \begin{array}{cccc}
    0 & w_{1 2} & \cdots & \ w_{1 N} \\
    w_{2 1} & 0 & \cdots & \ w_{2 N} \\
    \vdots & \vdots & \ddots & \vdots \\
    w_{N 1} & w_{N 2} & \cdots & 0
  \end{array}
  \right) \; .
\end{equation}
Notice that all the diagonal elements are identical to zero, $w_{i i} \equiv 0$. Self-loops (autapses) actually exist in some types of neurons and they could induce rich dynamical phenomena~\cite{Wang-Chen-2015}. In the present simplified model, we assume that the spontaneous relaxation term of Eq.~(\ref{eq:xevol}) has already incorporated the effect of possible self-loops.

In response to an external signal $\bm{s}$, the internal state of the linear recurrent dynamics (\ref{eq:linmod}) at time $t$ is
\begin{equation}
\label{eq:timeevol}
\bm{x}(t) = \bigl[\bm{I} - \textrm{e}^{-(\bm{I} + \bm{W}) t} \bigr]
( \bm{I} + \bm{W} )^{-1} \bm{s} \; ,
\end{equation}
where $\bm{I}$ is the identity matrix ($I_{i i}=1$ for the diagonal elements, $I_{i j}= 0$ for $i \neq j$). The net driving force of this response at time $t$ is simply ${\rm d} \bm{x}/{\rm d} t$, and it is linearly depending on $\bm{s}$,
\begin{equation}
    \label{eq:dxdt}
    \frac{ {\rm d} \bm{x}}{{\rm d} t} = \textrm{e}^{-(\bm{I} + \bm{W}) t} \bm{s} \; .
\end{equation}

Notice that, for $\bm{x}(t)$ to be convergent in this linear model, the real part of every eigenvalue of the composite matrix $(\bm{I} + \bm{W})$ needs to be positive. These eigenvalue conditions are checked in our numerical computations. (These constraints on the synaptic matrix may be unnecessary if the feedback functions are  nonlinear and bounded, such as $f_i( \bm{x}) = a_0 \tanh \bigl( \sum_{j \neq i} w_{i j} x^j \bigr)$ with $a_0$ being a positive constant.)

\subsection{Prediction and prediction error}

This single layer of neurons is tasked with processing many input signal vectors $\bm{s}_\alpha$, with index $\alpha = 1, 2, \ldots, P$. The total number $P$ of input samples is much larger than the total number $N$ of neurons. Each of these input vectors $\bm{s}_\alpha$ will be converted into an internal steady state $\bm{x}_\alpha = (x_\alpha^1, \ldots, x_\alpha^N)^\top$ by the recurrent dynamics (\ref{eq:xevol}), as
\begin{equation}
\bm{x}_\alpha = (\bm{I} + \bm{W})^{-1} \bm{s}_\alpha \; .
\label{eq:xss}
\end{equation}
The internal representation $\bm{x}_\alpha$ is a linear transformation of $\bm{s}_\alpha$. From this expression we see that the input signal vector $\bm{s}_\alpha$ has been decomposed into two parts,
\begin{equation}
\label{eq:sdecomp}
\bm{s}_\alpha = \bm{W} \bm{x}_\alpha + \bm{x}_\alpha =
\bm{p}_\alpha + \bm{x}_\alpha \; .
\end{equation}

The $i$-th element of the first vector $\bm{p}_\alpha \equiv \bm{W} \bm{x}_\alpha$ is $p_\alpha^i = \sum_{j \neq i} w_{ i j} x_\alpha^j$ and it is independent of $x_\alpha^i$. We can therefore interpret $p_\alpha^i$ as the prediction by the other neurons concerning the input signal $s_\alpha^i$ of neuron $i$. Each neuron $j$ contributes a term $w_{i j} x_\alpha^j$ to the predicted input signal at neuron $i$, and the column vector $(w_{1, j}, \ldots, w_{j-1, j}, 0, w_{j+1, j}, \ldots, w_{N, j})^\top$ characterizes the predictive role of neuron $j$. We refer to $\bm{p}_\alpha$ as the prediction vector. Then, Eq.~(\ref{eq:sdecomp}) indicates that the internal state $x_\alpha^i$ of neuron $i$ is also serving as a  prediction error. When  $\bm{x}_\alpha$ is received as input by another layer of neurons, it contains the residual properties of the signal vector $\bm{s}_\alpha$ that has not yet been predicted by the single-layer internal model $\bm{W}$. In other words, $x_\alpha^i$ is both an internal state of neuron $i$ itself and a prediction error for $s_\alpha^i$. This dual role might be of real biological significance,  as it relieves the necessity of introducing extra neurons for the sole purpose of computing prediction error~\cite{Aguilera-etal-2022}. In the literature, special error-computing neurons are commonly employed in hierarchical predictive coding network models~\cite{Huang-Rao-2011}. Such neurons may not be strictly necessary (and indeed the biological evidence in support of their existence is weak). Top-down predictive messages can be directly fed into the single-layer response dynamics. The simplest way is to add a term $-\bm{h}(t)$ into the right-hand side of Eq.~(\ref{eq:xevol}), with $\bm{h}(t)$ being the higher-level prediction concerning the input $\bm{s}$.

\begin{figure}
\centering
\includegraphics[width=0.7\linewidth]{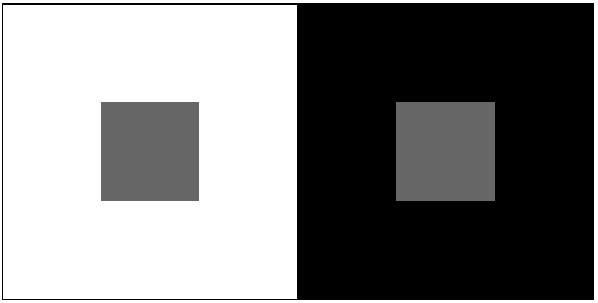}
\caption{
An example of perceptive illusion. The gray intensities of the $18$ small square blocks are: $0.6$ for the two central squares; $0.0$ (white) for the eight squares surrounding the left central square; $1.0$ (black) for the eight squares surrounding the right central square. The synaptic weights from a square $j$ to all its surrounding squares $i$ are set to  $w_{i j} = 0.05$ and all the other synaptic weights are set to be zero. The perceived gray intensities of the left and right central squares are $0.624$ and $0.254$, respectively.
}
\label{fig:shadecontrast}
\end{figure}

The decomposition (\ref{eq:sdecomp}) will cause an interesting phenomenon of perceptive illusion. We explain this by a simple example, the gray image shown in Fig.~\ref{fig:shadecontrast}. The eight small white squares surrounding the left central small square have gray intensity $0.0$, the eight small black squares surrounding the right central small square have gray intensity $1.0$, and the two central small squares have identical gray intensity $0.6$. Let us set the weight $w_{i j}$ from a square $j$ to all its nearest surrounding squares $i$ to be $w_{i j} = 0.05$ and set all the other weights to be zero. Then according to Eq.~(\ref{eq:xss}), the output gray value of the left small square will be $0.624$ and that of the right small square is $0.254$. The linear predictive coding model ``perceives" the left central square to be much more darker than the right central square, even though their actual gray intensity is the same. This is an unconscious predictive perception, and this outcome is consistent with the conscious visual perception of humans. Unconscious predictive coding may be  contributing to conscious optical illusions. We notice that recent  computational investigations suggested that more complex visual illusions, such as the famous Kanizsa contours~\cite{Kanizsa-1976}, could be quantitatively explained by hierarchical predictive coding models with feedback interactions~\cite{Pang-etal-2021}.

\section{Prediction error minimization}
\label{sec:adapt}

\subsection{Mean squared prediction error}

For the linear predictive coding model with $P$ input signal vectors $\bm{s}_\alpha$, we define the mean squared prediction error $\varepsilon$ as
\begin{equation}
\varepsilon = \frac{1}{2 P} \sum\limits_{\alpha=1}^P \bigl( \bm{x}_\alpha \bigr)^2
=  \frac{1}{2 P} \sum\limits_{\alpha=1}^P \sum\limits_{i=1}^N \bigl( x_\alpha^i )^2 \; ,
\label{eq:varepsilon}
\end{equation}
where $\bm{x}_\alpha$ is the prediction error vector corresponding to input $\bm{s}_\alpha$. As $x_\alpha^i$ is also the internal state of neuron $i$, the mean squared prediction error can also be understood as the mean metabolic energy needed to represent an external input. The synaptic weights $w_{i j}$ of lateral interactions are modifiable on time scales much longer than that of the response dynamics (\ref{eq:xevol}). It is natural for us to assume that these synaptic weights will slowly adapt to the inputs to minimize the mean squared prediction error $\varepsilon$.

Here we are interested in the properties of the optimized synaptic weights. To better appreciate the essence of this optimization task, we now rewrite the expression of $\varepsilon$ in an alternative form. The empirical correlation matrix $\bm{A}$ for the $N$ neurons is defined as $\bm{A} \equiv (1/P) \sum_{\alpha=1}^P \bm{s}_\alpha \bm{s}_\alpha^\top$, with elements
\begin{equation}
    A_{i j} = \frac{1}{P}
    \sum\limits_{\alpha = 1}^{P} s_\alpha^i s_\alpha^j \; .
    \label{eq:Aije}
\end{equation}
This real symmetric matrix has $N$ non-negative eigenvalues $\lambda_m$, ranked in descending order $\lambda_1 \geq \lambda_2 \geq \ldots \geq \lambda_N$. The corresponding eigenvectors are denoted as $\bm{u}_m \equiv (u_{1 m}, u_{2 m}, \ldots, u_{N m})^\top$, which satisfy the orthonormal property $\bm{u}_m^\top \bm{u}_m = 1$ and $\bm{u}_m^\top \bm{u}_n = 0$ for $m \neq n$. Any input signal vector $\bm{s}_\alpha$ could be uniquely expanded as
\begin{equation}
\label{eq:spca}
  \bm{s}_\alpha = \sum\limits_{m=1}^{N} c_{\alpha}^m \bm{u}_m \; ,
\end{equation}
where $c_{\alpha}^m$ is the coordinate along the $m$-th eigenvector, namely $c_\alpha^m = \bm{u}_m^\top \bm{s}_\alpha$. The moments of these coordinates have the following important properties:
\begin{eqnarray}
  \bigl\langle ( c_{\alpha}^m )^2 \bigr\rangle & \equiv &
  \frac{1}{P} \sum\limits_{\alpha}
  ( c_{\alpha}^m )^2 = \lambda_m \; , 
  \label{eq:cici}\\
  \bigl\langle  c_{\alpha}^m c_{\alpha}^n \bigr\rangle & \equiv &
  \frac{1}{N} \sum\limits_{\alpha}
  c_{\alpha}^m c_{\alpha}^n  = 0  \; ,  \quad\quad (m \neq n ) \; .
  \label{eq:cicj}
\end{eqnarray}
Let us denote by $\bm{z}_m = (z_{1 m}, \ldots, z_{N m})^\top$ the internal representation of the eigenvector $\bm{u}_m$,
\begin{equation}
\bm{z}_m = (\bm{I} + \bm{W} )^{-1} \bm{u}_m \; .
\end{equation}
In matrix form, this means
\begin{eqnarray}
  & & 
  \left(
  \begin{array}{cccc}
    1 & w_{1 2} & \cdots & w_{1 N} \\
    w_{2 1} & 1 & \cdots & w_{2 N} \\
    \vdots & \vdots & \ddots & \vdots \\
    w_{N 1} & w_{N 2} & \cdots & 1
  \end{array}
  \right) 
  \left(
  \begin{array}{cccc}
    z_{1 1} & z_{1 2} & \cdots & z_{1 N} \\
    z_{2 1} & z_{2 2} & \cdots & z_{2 N} \\
    \vdots & \vdots & \ddots & \vdots \\
    z_{N 1} & z_{N 2} & \cdots & z_{N N}
  \end{array}
  \right) 
  \nonumber \\
  & & \quad \quad 
  =
  \left(
  \begin{array}{cccc}
    u_{1 1} & u_{1 2} &
    \cdots & u_{1 N} \\
    u_{2 1} & u_{2 2} &
    \cdots & u_{2 N} \\
    \vdots & \vdots & \ddots & \vdots \\
    u_{N 1} & u_{N 2} &
    \cdots & u_{N N}
  \end{array}
  \right) \; .
\label{eq:lineareq}
\end{eqnarray}
For any input signal vector $\bm{s}_\alpha$, because of Eq.~(\ref{eq:spca}), we know that its internal representation is
\begin{equation}
\bm{x}_\alpha \equiv (\bm{I} + \bm{W} )^{-1} \bm{s}_\alpha 
= \sum\limits_{m = 1}^N c_\alpha^m \bm{z}_m \; .
\label{eq:xexpan}
\end{equation}
This expression has the same form as Eq.~(\ref{eq:spca}), but notice that the basis vectors $\bm{z}_m$ are not necessarily orthogonal to each other and also that the  squared norm $\| \bm{z} \|^2 \equiv \bm{z}_m^\top \bm{z}_m \neq 1$. The coordinates $c_\alpha^m$ are fixed by the input dataset, and the optimization targets are then the $N$ internal vectors $\bm{z}_m$. The mean squared prediction error (\ref{eq:varepsilon}) is 
\begin{eqnarray}
\varepsilon 
& = & 
\frac{1}{2 P} \sum\limits_{\alpha=1}^{P} \sum\limits_{m = 1}^N \sum\limits_{n = 1}^{N} c_\alpha^m c_\alpha^n  \bm{z}_m^\top \bm{z}_n
\nonumber \\
& = & \frac{1}{2} \sum\limits_{m = 1}^{N} \lambda_m ( \bm{z}_m )^2
\label{eq:prex1}
\\
 & = & \frac{1}{2} \sum\limits_{m = 1}^{N} \lambda_m \bigl[ (\bm{I}+ \bm{W}) ^{-1} \bm{u}_m \bigr]^2 \; .
\label{eq:prex2}
\end{eqnarray}
In deriving Eq.~(\ref{eq:prex1})  we have used the properties (\ref{eq:cici}) and (\ref{eq:cicj}). The summation in Eq.~(\ref{eq:prex1}) does not contain inner product terms $\bm{z}_m^\top \bm{z}_n$ with $m \neq n$. This fact means that the orthogonality of the internal vectors $\bm{z}_m$ is not a necessary condition for the minimization of $\varepsilon$.

The two equivalent expressions (\ref{eq:prex1}) and (\ref{eq:prex2}) reveal that the mean squared prediction error does not depend on the details of the $P$ input signal vectors but only on the eigenvalues and eigenvectors of the correlation matrix $\bm{A}$. For the leading (largest) eigenvalues $\lambda_m$ ($m=1, 2, \ldots$), it is desirable to reduce the squared norm of the corresponding internal vectors $\bm{z}_m$. On the other hand, there is no much necessity to optimize an internal vector $\bm{z}_n$ if the eigenvalue $\lambda_n$ is close to zero. The first eigenvector $\bm{u}_1$ and eigenvalue $\lambda_1$ may be strongly related to the mean vector $(1/P) \sum_\alpha \bm{s}_\alpha$ of the input signals. The remaining eigenvectors and eigenvalues are mainly related to co-variance of the input signal vectors. We expect that the leading terms $\lambda_m (\bm{z}_m)^2$ of Eq.~(\ref{eq:prex1}) with $m \geq 2$ will be roughly  equal, i.e., $\| \bm{z}_m \| \sim \lambda_m^{-1/2}$ for $m=2, 3, \ldots$ as a result of optimization. Because $c_\alpha^m \sim \lambda_m^{1/2}$ according to Eq.~(\ref{eq:cici}), the projections of the internal state $\bm{x}_\alpha$ on the different directions $\hat{\bm{z}}_m \equiv  \bm{z}_m / \| \bm{z}_m \|$ of $m = 2, 3, \ldots$ will be comparable in magnitude. Suppose only a few number (say $M\approx 100$) of $\lambda_m$ values are important.  Then, according to Eq.~(\ref{eq:xexpan}), the internal representation $\bm{x}_\alpha$ will be
\begin{equation}
    \bm{x}_\alpha  \approx \tilde{c}_\alpha^1 \hat{\bm{z}}_1 + 
    \sum\limits_{m=2}^M \tilde{c}_\alpha^m \hat{\bm{z}}_m \; ,
    \label{eq:xapprox}
\end{equation}
where $\tilde{c}_\alpha^m \equiv c_\alpha^m \| \bm{z}_m \|$. The magnitudes of the coefficients $\tilde{c}_\alpha^m$ for $2 \leq m \leq M$ will be roughly equal if the scaling property $\lambda_m^{1/2} \| \bm{z}_m \| \sim 1$ is valid. The $N$ internal direction vectors $\hat{\bm{z}}_m$ may not be strictly orthogonal to each other, but instead the angles between them may slightly deviate from $\pi/2$.

\subsection{Evolution of synaptic weights}
\label{sec:tcost}

We minimize the mean squared prediction error (\ref{eq:prex2}) by the method of gradient descent, under the constraint that the real parts of all the eigenvalues of $(\bm{I}+ \bm{W})$ being positive. In addition, considering that maintaining a nonzero synaptic weight has a metabolic cost, we introduce a quadratic energy term to each synaptic weight. The total cost function of the minimization problem is then
\begin{equation}
  C( \bm{W} ) = \frac{1}{2} \sum\limits_{m = 1}^{N}
  \lambda_m \bigl[ (\bm{I} + \bm{W})^{-1} \bm{u}_m \bigr]^2 + 
  \frac{\eta}{2 N} \sum\limits_{i, j} \bigl( w_{i j} \bigr)^2 \; ,
\end{equation}
where $\eta$ is an adjustable penalty parameter (the scaling factor $N^{-1}$ ensures that the two summation terms in the above expression are of the same order, that is, proportional to $N$).

The first derivative of this cost function with respect to synaptic weight $w_{i j}$ is
\begin{eqnarray}
\frac{ \partial C}{\partial w_{i j}} & = &  - \sum\limits_{m, n, p}
(\bm{I}+\bm{W})^{-1}_{j m} A_{m n} 
(\bm{I}+\bm{W})^{-1}_{p n} (\bm{I}+\bm{W})^{-1}_{p i}
\nonumber \\
& & \quad  + \frac{\eta}{N} w_{i j} \; .
\label{eq:slopeC}
\end{eqnarray}
In deriving this expression, we have used the following two relations
\begin{eqnarray}
    \Bigl(
    \frac{\partial (\bm{I} + \bm{W})^{-1}}{\partial w_{i j} }
    \Bigr)_{m n} & = & 
    - (\bm{I}+ \bm{W})^{-1}_{j n} (\bm{I} + \bm{W})^{-1}_{m i} \; , \\
    A_{m n} & = &
    \sum_{k} \lambda_k u_{m k} u_{n k} \; .
\end{eqnarray}
To minimize the total cost $C$ by gradient descent, we modify all the synaptic weights $w_{i j}$ ($i \neq j$) simultaneously according to
\begin{equation}
    w_{i j} \ \leftarrow \ w_{i j} - \gamma \frac{ \partial C}{\partial w_{i j}} \; ,
    \label{eq:wgdevol}
\end{equation}
where $\gamma$ is a small learning rate.

From the expression (\ref{eq:slopeC}) of cost gradients, we observe that
\begin{equation}
\frac{ \partial C}{\partial w_{i j}} \neq 
\frac{ \partial C}{\partial w_{j i} } \; ,
\end{equation}
although the correlation matrix $\bm{A}$ is symmetric. Then the adaptation of $w_{i j}$ and $w_{j i}$ following (\ref{eq:wgdevol}) and starting from $w_{i j}=w_{j i}=0$ will lead to breaking of symmetry between these two synaptic weights, that is, $w_{i j} \neq w_{j i}$. We have checked by exact computation that this spontaneous symmetry-breaking phenomenon occurs even if there are only two neurons, $N=2$.

\section{Numerical results}
\label{sec:mnist}

We apply the lateral predictive coding model to a widely used real-world dataset, the MNIST dataset of hand-written digits~\cite{LeCun-etal-1998}, with the purpose of gaining some empirical insights on the effects of lateral recurrent interactions. There are $P=60000$ gray images of $28 \times 28$ pixels for the ten digits, each of which serves as an input vector ($\bm{s}_\alpha$). We attach a neuron to each of the $N=784$ pixels, and neurons and pixels will be mentioned interchangably in this section. The original pixel values are integers ranging from $0$ to $255$. Here we linearly re-scale these values to the range $[0, 1]$. The mean input vector, denoted as $\overline{\bm{s}} \equiv  \sum_{\alpha=1}^P \bm{s}_\alpha / P$, is a positive vector. The mean prediction vector and the mean prediction error vector are denoted by $\overline{\bm{p}}$ and $\overline{\bm{x}}$, respectively. Naturally, these three mean vectors satisfy the relation $\overline{\bm{s}} = \overline{\bm{p}} + \overline{\bm{x}}$.

For convenience of later discussions, we define the (cosine) similarity $q(\bm{v}, \bm{y})$ of two generic $m$-dimensional vectors $\bm{v} = (v_1, \ldots, v_m)^\top$ and $\bm{y}=(y_1, \ldots, y_m)^\top$ as
\begin{equation}
\label{eq:simdef}
q( \bm{v}, \bm{y} ) \equiv 
\frac{ \bm{v}^\top \bm{y} }{\| \bm{v} \| \ \| \bm{y} \|} 
 =  \frac{ \sum_{k} v_k y_k}{ \bigl[\sum_{i} v_i^2\bigr]^{\frac{1}{2}} \bigl[ \sum_{j} y_j^2\bigr]^{\frac{1}{2}}}\; .
\end{equation}
This similarity index measures the angle between $\bm{v}$ and $\bm{y}$. For example, if $\bm{v}$ and $\bm{y}$ point to the same direction, then $q(\bm{v}, \bm{y}) = 1$; if they are orthogonal to each other, then $q(\bm{v}, \bm{y}) = 0$. 

\begin{figure*}
\centering
\subfigure[]{
  \includegraphics[width=0.26\linewidth]{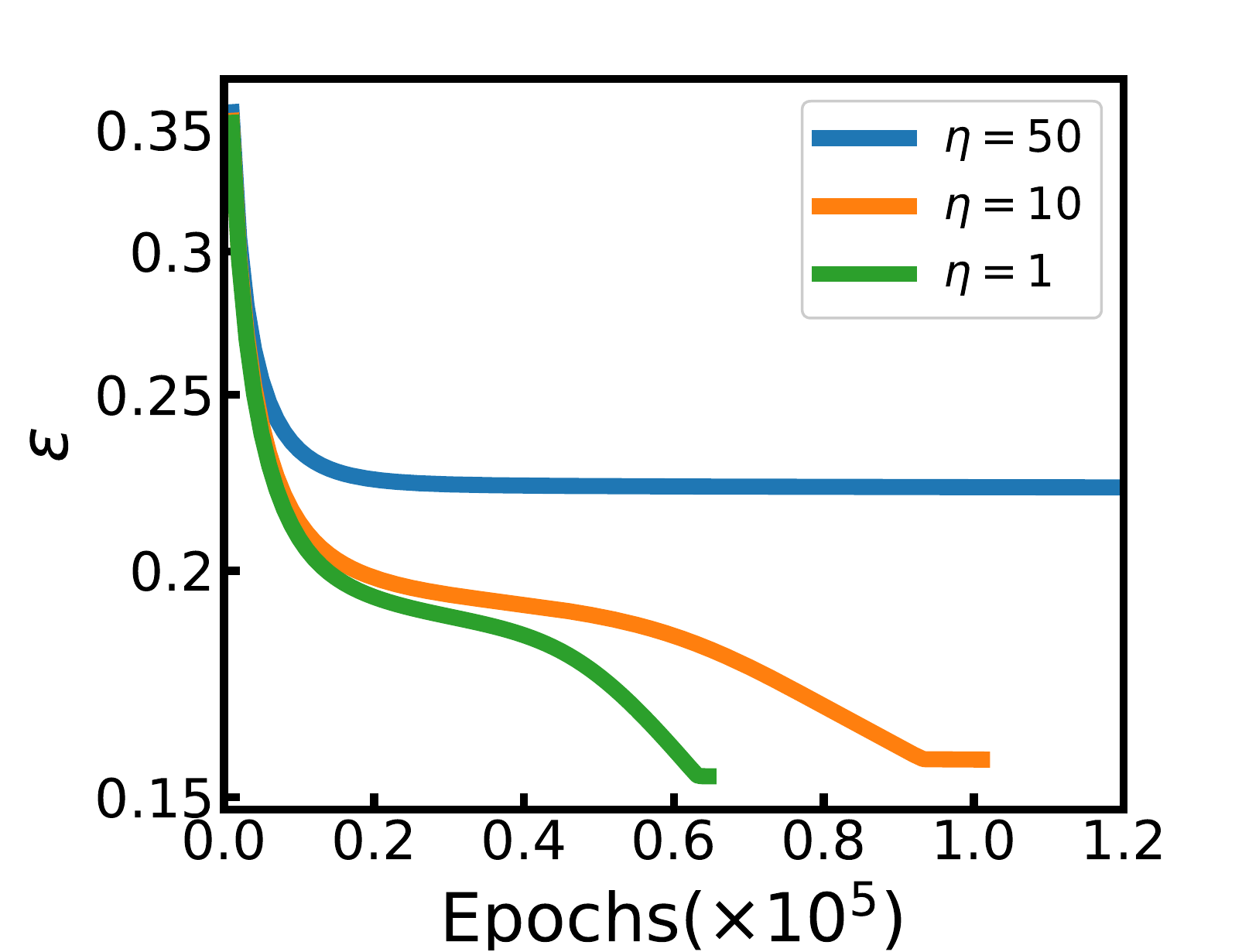}
\label{fig:varepsilon}
}
\subfigure[]{
  \includegraphics[width=0.25\linewidth]{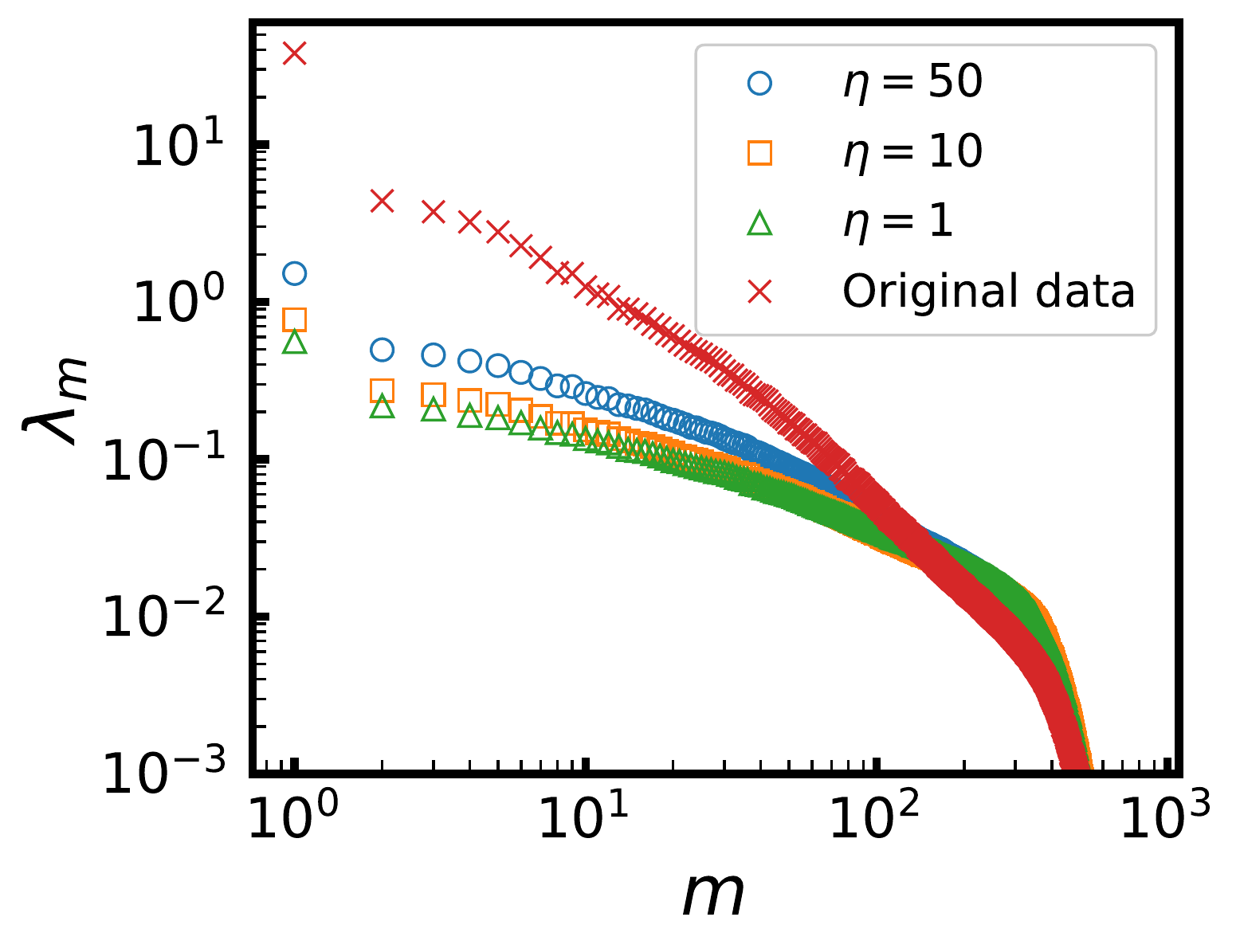}
\label{fig:lambdaSX}
}
\subfigure[]{
  \includegraphics[width=0.25\linewidth]{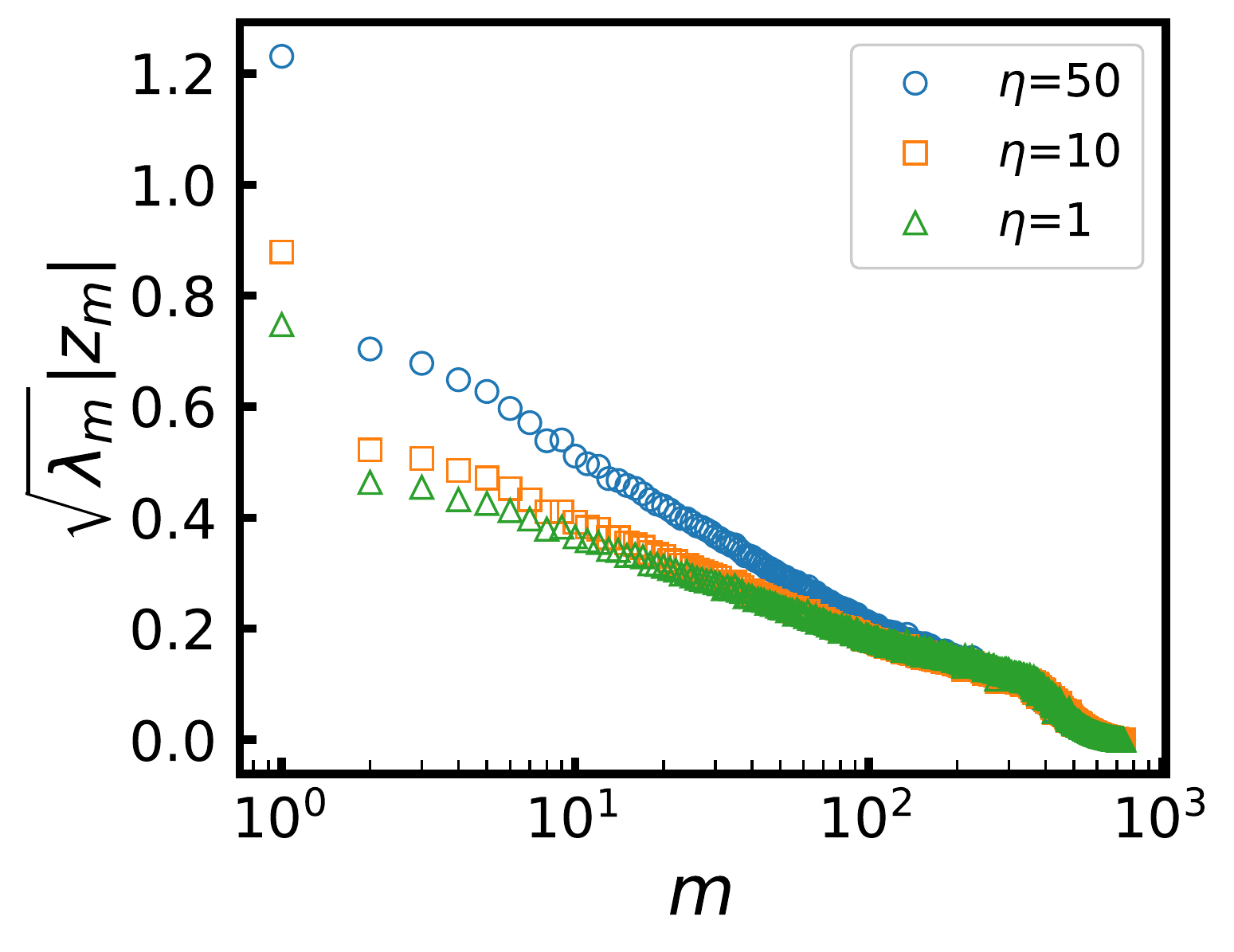}
\label{fig:lambdazk}
}
\caption{
(a) Evolution of the mean squared prediction error $\varepsilon$ (relative to the initial value before weight adaptation). (b) The eigenvalues $\lambda_m$ of the correlation matrix of the input data samples $\bm{s}_\alpha$ and of the correlation matrix of the internal states $\bm{x}_\alpha$. (c) The values of $\lambda_m^{1/2} \| \bm{z}_m \|$ where $\lambda_m$ is an eigenvalue of the correlation matrix of the input data samples. The weight penalty parameter is $\eta = 50$, $10$, or $1$.
}
\label{fig:lambda}
\end{figure*}

\subsection{Learning procedure}

The linear predictive coding model requires all the eigenvalues of the composite matrix $(\bm{I}+\bm{W})$ to have positive real part, so we numerically check all the eigenvalues of this matrix  every $T$ iteration steps (epochs) during the evolution process of Eq.~(\ref{eq:wgdevol}).  The inspection interval is initially set to be $T=1000$. After the eigenvalue constraints are checked to be  violated for the first time, the value of $T$ is reduced to $T=100$ and then fixed to this value. Each time the matrix $(\bm{I}+\bm{W})$ is checked to have at least one eigenvalue with negative real part, the learning rate is reduced by half ($\gamma \leftarrow \gamma/2$), and  the synaptic matrix $\bm{W}$ is also properly reset to carry out the next $T$ evolution epochs. One biologically plausible way of resetting the synaptic weight matrix is by homeostatic scaling-down of all its elements by the same ratio~\cite{Diering-etal-2017}. Another simple way is to simply reset the weight matrix to the matrix $\bm{W}$ that was attained and recorded $T$ epochs earlier.  The numerical results reported in this section were obtained by the second resetting method, but we have checked that the final numerical values of the synaptic weights are not sensitive to the particular method used to guarantee the eigenvalue condition, nor to whether the iteration (\ref{eq:wgdevol}) was performed synchronously or in random sequential order. The learning rate is initially set to be $\gamma = 0.001$. We train the network using PyTorch (version 1.10.0) and Python (version 3.9.7), which are quite convenient for matrix manipulations. 

We consider three representative values for the penalty parameter: strong penalty, $\eta = 50$; moderate penalty, $\eta = 10$; weak penalty, $\eta = 1$. For $\eta = 50$ we find that all the eigenvalues of $(\bm{I}+\bm{W})$ never violate the positivity  condition during the whole evolution process, while weight matrix resettings are needed at $\eta = 10$ and $\eta=1$. We find that the properties of the systems obtained at different values of $\eta$ are actually very similar qualitatively.

\begin{figure}
\centering
\subfigure[]{
  \includegraphics[width=0.465\linewidth]{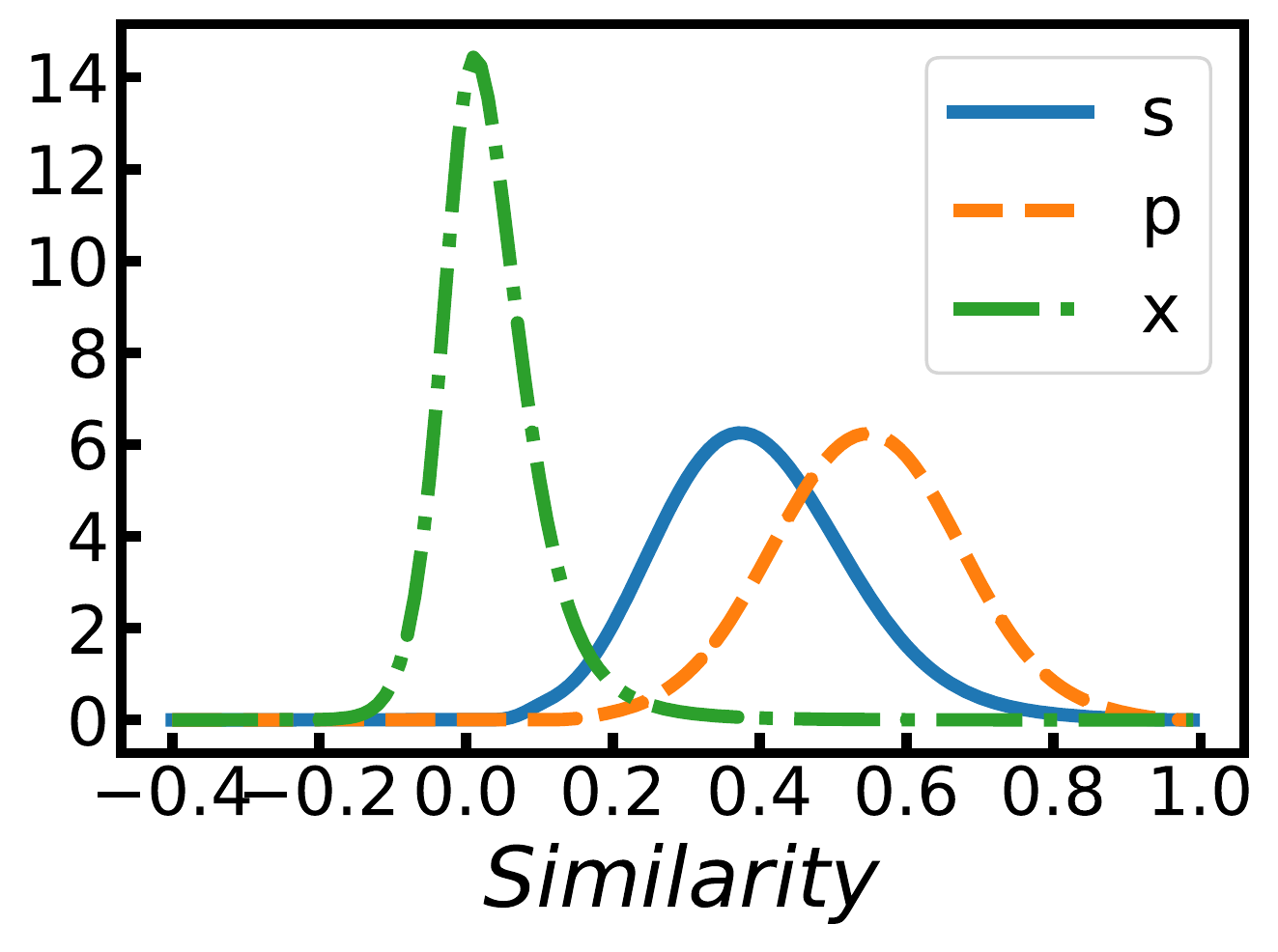}
\label{fig:dataqval:A}
}
\subfigure[]{
  \includegraphics[width=0.465\linewidth]{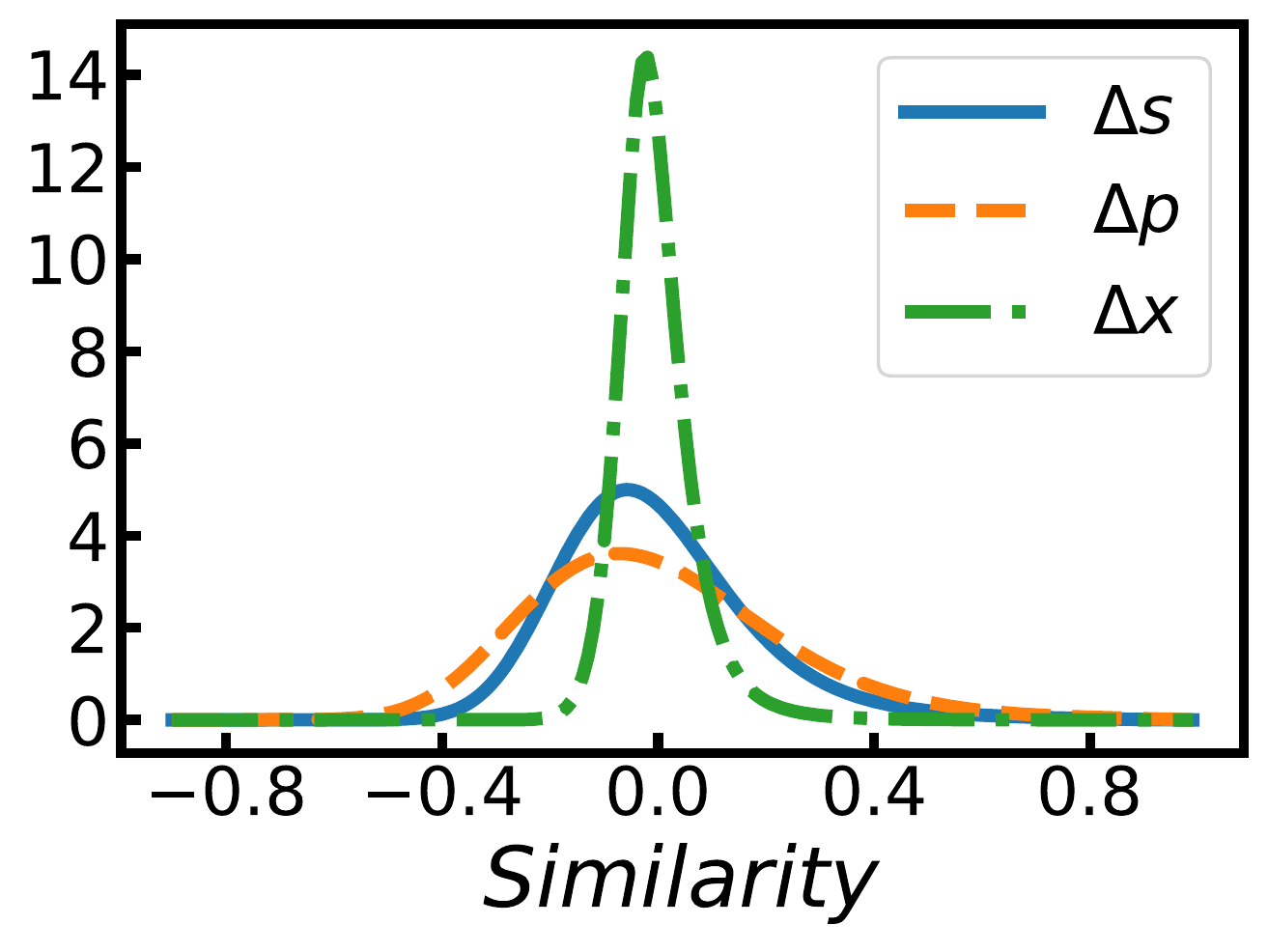}
\label{fig:dataqval:B}
}
\caption{
Histogram of similarity values between $60000$ vectors. Solid curves: input signal vectors $\bm{s}_\alpha$ (a) and the mean-corrected ones $\Delta \bm{s}_\alpha = \bm{s}_\alpha -  \overline{\bm{s}}$ (b); dashed curves: prediction vectors $\bm{p}_\alpha$ (a) and the mean-corrected ones $\Delta \bm{p}_\alpha = \bm{p}_\alpha -  \overline{\bm{p}}$ (b); dot-dashed curves: prediction error vectors $\bm{x}_\alpha$ (a) and the mean-corrected ones $\Delta \bm{x}_\alpha = \bm{x}_\alpha -  \overline{\bm{x}}$ (b). The weight penalty is $\eta=1$.
}
\label{fig:dataqval}
\end{figure}

Figure~\ref{fig:varepsilon} shows the decay curves of the mean squared prediction error $\varepsilon$ with the learning epoch. At large weight penalty ($\eta = 50$) the value of $\varepsilon$ decreases to a final value which is about $0.23$ of the initial value in about $10^4$ epochs. At moderate or low weight penalty ($\eta = 10$ or $\eta=1$) it takes longer for $\varepsilon$ to saturate, but the final value is considerably lower ($\approx 0.16$ of the initial value). Resetting or rescaling of the weight matrix is needed at $\eta = 10$ or $\eta = 1$. The first $500$ or so eigenvalues of the correlation matrix $\bm{A}$ [Eq.~(\ref{eq:Aije})] of the input vectors $\bm{s}_\alpha$ and the corresponding eigenvalues of the correlation matrix of the internal states $\bm{x}_\alpha$ are shown in Fig.~\ref{fig:lambdaSX} for comparison. The first eigenvalue $\lambda_1$ of the correlation matrix is much larger than the second eigenvalue $\lambda_2$. This is mainly caused by the fact that the mean input vector $\overline{\bm{s}}$ is a relatively large positive vector.  We see that, as a consequence of the decomposition (\ref{eq:sdecomp}), the first $100$ eigenvalues of the internal states are considerably reduced as compared with those of the input data. The relationship $\| \bm{z}_m \| \sim \lambda_m^{-1/2}$ is confirmed to be roughly true for $m\geq 2$ [Fig.~\ref{fig:lambdazk}]. 

The distribution of similarity $q(\bm{s}_\alpha, \bm{s}_\beta)$ between two input images of the MNIST dataset, and the corresponding distributions of similarity $q(\bm{p}_\alpha, \bm{p}_\beta)$ and $q(\bm{x}_\alpha, \bm{x}_\beta)$ are compared in Fig.~\ref{fig:dataqval:A}. The distribution of $q(\bm{x}_\alpha, \bm{x}_\beta)$ is sharply peaked around zero, suggesting that the prediction error vectors $\bm{x}_\alpha$ of the $60000$ data samples are approximately orthogonal to each other. This is a clear demonstration of redundancy reduction in $\bm{x}_\alpha$. In comparison,  we find that both $q(\bm{s}_\alpha, \bm{s}_\beta)$ and $q(\bm{p}_\alpha, \bm{p}_\beta)$ are peaked at relatively large positive values, which may be mainly due to the fact that the mean input vector $\overline{\bm{s}}$ and the mean prediction vector $\overline{\bm{p}}$ are both non-zero. If we subtract the mean vectors to get the mean-corrected vectors ($\Delta \bm{s}_\alpha = \bm{s}_\alpha - \overline{\bm{s}}$, $\Delta \bm{p}_\alpha = \bm{p}_\alpha - \overline{\bm{p}}$, and $\Delta \bm{x}_\alpha = \bm{x}_\alpha - \overline{\bm{x}}$), the similarity distributions of $\Delta \bm{s}_\alpha$ and $\Delta \bm{p}_\alpha$ both are shifted to be peaked close to zero and also their standard deviations become slightly more broader [Fig.~\ref{fig:dataqval:B}]. On the other hand, this mean-correction treatment does not have significant effect on the prediction error,  probably because the mean vector $\overline{\bm{x}}$ are already quite small (the mean value of its $N$ elements is about $0.01$ at $\eta=1$).

\subsection{Nonsymmetry and sparsity of synaptic weights}

\begin{figure*}
\centering
\subfigure[]{
  \includegraphics[width=0.31\linewidth]{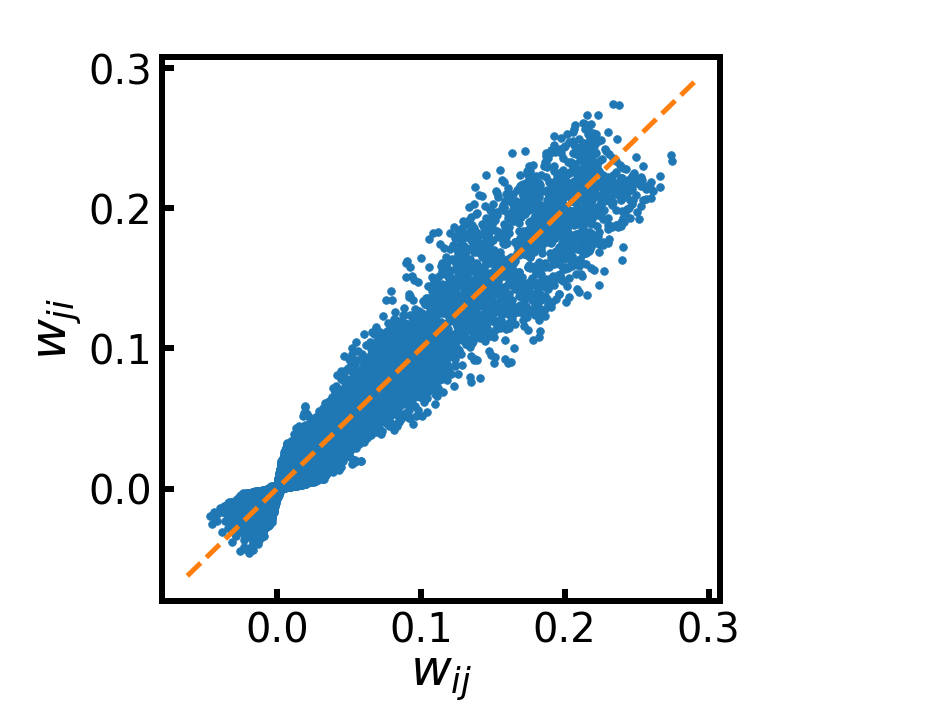}
  \label{fig:symmetry:A}
}
\subfigure[]{
  \includegraphics[width=0.31\linewidth]{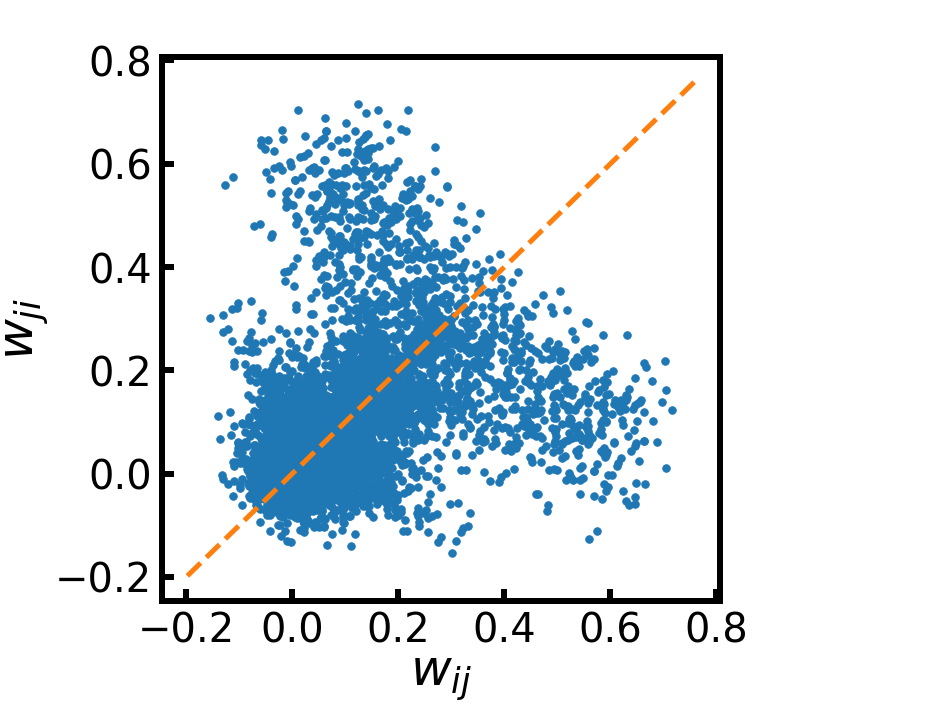}
  \label{fig:symmetry:B}
}
\subfigure[]{
 \includegraphics[width=0.31\linewidth]{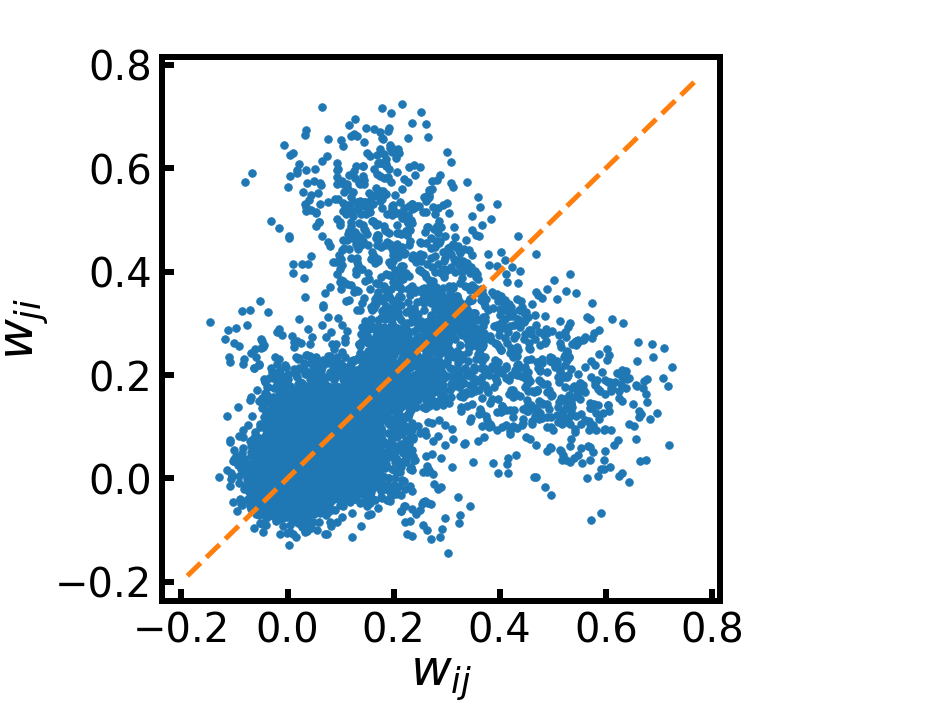}
  \label{fig:symmetry:C}
}
\caption{
\label{fig:symmetry}
Symmetry breaking of the synaptic weight matrix $\bm{W}$. Here $w_{i j}$ is the synaptic weight from neuron $j$ to neuron $i$. The penalty parameter is $\eta = 50$ (a), $\eta = 10$ (b),  and $\eta = 1$ (c).
}
\end{figure*}

\begin{figure}
\centering
\subfigure[]{
  \includegraphics[width=0.95\linewidth]{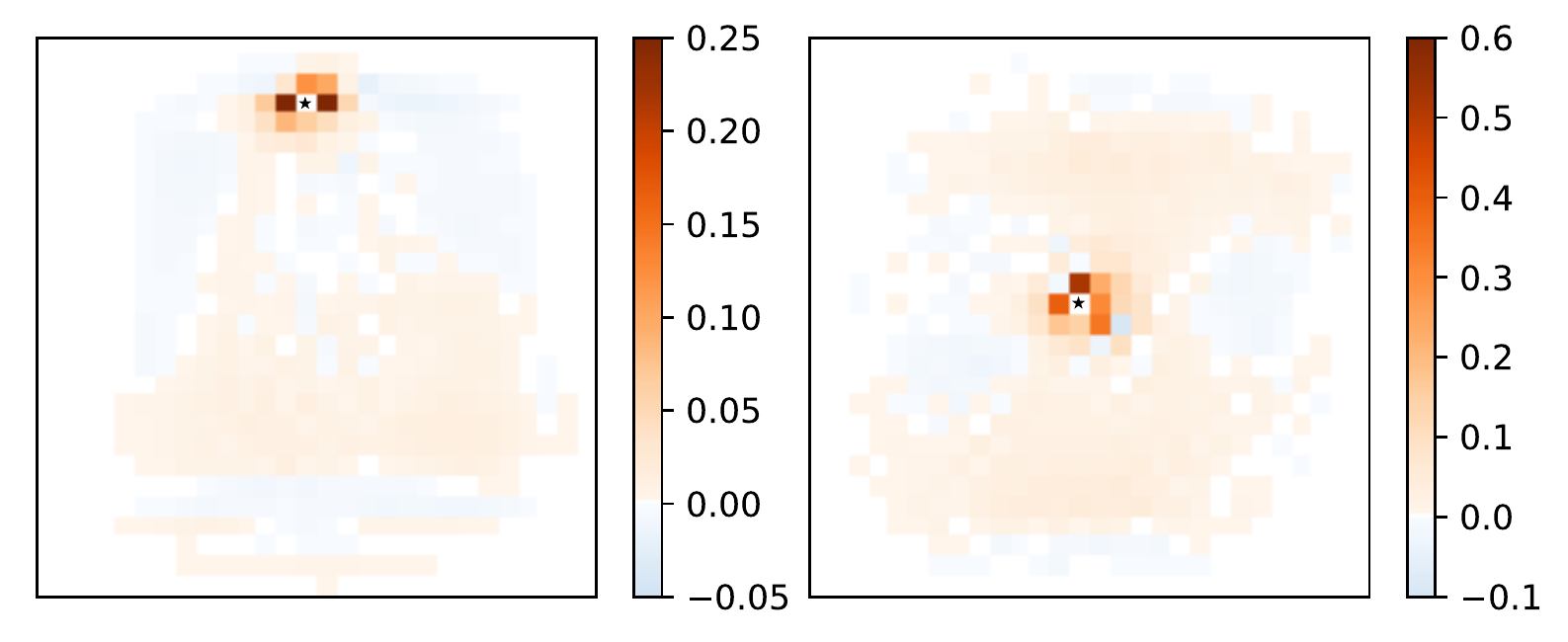}
  \label{fig:winput}
}
\subfigure[]{
  \includegraphics[width=0.95\linewidth]{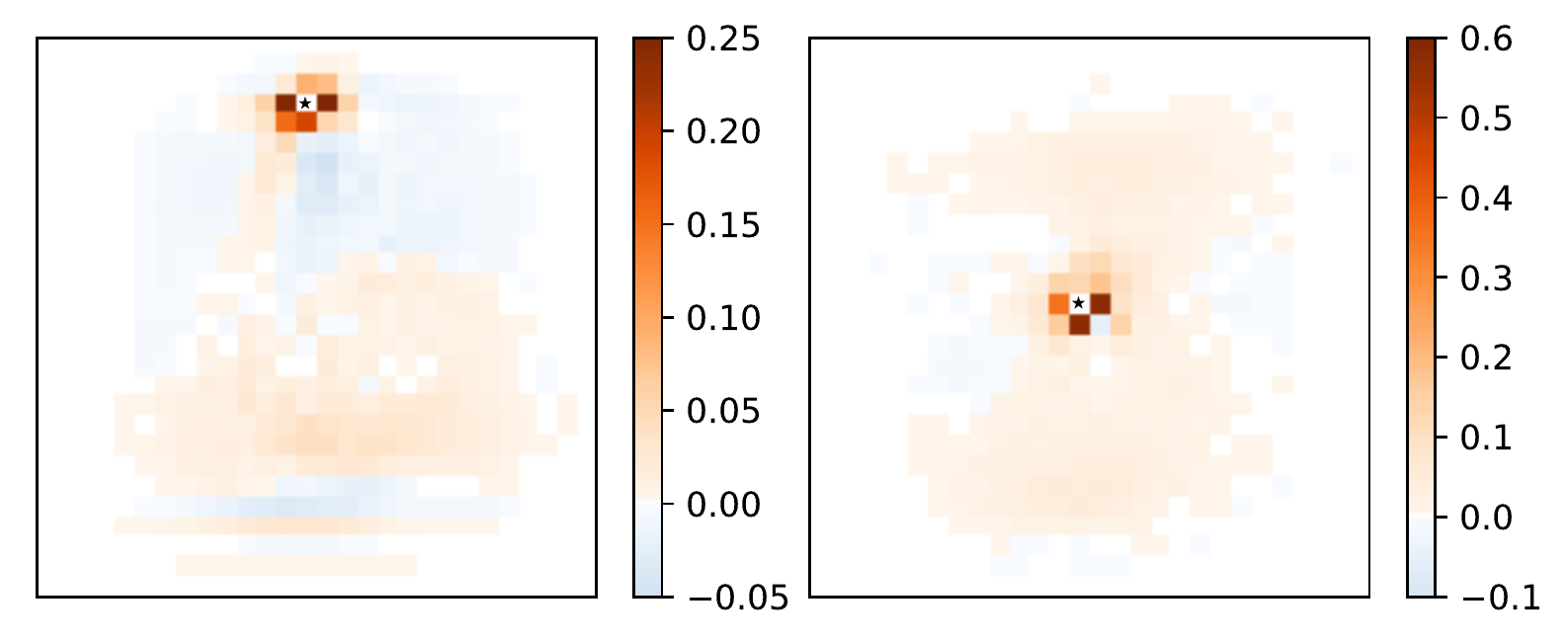}
  \label{fig:woutput}
}
\caption{Receptive fields $w_{i j}$ and projection fields $w_{j i}$ for two neurons $i$, whose positions at $(4, 14)$ and $(14, 14)$ are marked by the `$*$' symbols. The weight penalty parameter is  $\eta = 1$. (a) The input synaptic weights $w_{i j}$ to the focal neuron $i$ from all the other neurons $j$. (b) The  output synaptic weights $w_{j i}$ from the focal neuron $i$ to all the other neurons $j$.
}
\label{fig:weightio}
\end{figure}

The gradient-descent dynamics starts from the all-zero synaptic weight matrix $\bm{W}$ ($w_{i j} = w_{j i} = 0$ for all the pair-wise interactions). Our simulation results confirm the theoretical expectation of Sec.~\ref{sec:tcost} that the symmetry property of $\bm{W}$ breaks down as a result of optimisation (Fig.~\ref{fig:symmetry}), and $w_{i j} \neq w_{j i}$. This nonsymmetry could be quite large for some of the neuron pairs. As an example, consider a neuron $i$ located at the central pixel $(14, 14)$ in Fig.~\ref{fig:weightio} and a neighboring neuron $j$ located at pixel $(15, 14)$. These two neurons are highly correlated in MNIST, with $A_{i j} = 0.304$. At $\eta = 1$ we find that $w_{i j}=0.145$ while $w_{j i} = 0.570$, which means that the state of neuron $i$ has a strong direct effect on that of neuron $j$ but the opposite is not true.

We quantify the average degree of nonsymmetry by the following ratio $\kappa$
\begin{equation}
\label{eq:kappa}
    \kappa  = \frac{1}{N (N-1)}
    \sum\limits_{i \neq j} 
    \frac{  | w_{i j} - w_{j i}| }{ \bigl( | w_{i j} | + | w_{j i}|
    \bigr)  } \; .
\end{equation}
At strong synaptic penalty ($\eta = 50$) the nonsymmetry ratio is relatively small ($\kappa = 0.278$); as the synaptic penalty is lowered to $\eta = 10$, the nonsymmetry ratio increases to a relatively large value of $\kappa = 0.514$; further decreasing the penalty to $\eta = 1$ only has a tiny effect on the nonsymmetry ratio ($\kappa = 0.476$). As the penalty value $\eta$ decreases, the synaptic weights have more flexibility to take larger values. Figure~\ref{fig:symmetry:B} and \ref{fig:symmetry:C} clearly demonstrate that, if the synaptic weight $w_{i j}$ from neuron $j$ to neuron $i$ is large enough ($w_{i j} > 0.3$), the reverse synaptic weight $w_{j i}$ from $i$ to $j$ is highly likely to be much smaller, with the sum $w_{i j} + w_{j i}$ being roughly a constant value. 

We define the lateral receptive field of a neuron $i$ as the subset of other neurons $j$ with their synaptic weights $w_{i j}$ to neuron $i$ significantly deviating from zero. For the two-dimensional MNIST system, we find that the receptive field of each neuron $i$ is considerably localized and is sparse: only a few of the input  synaptic weights $w_{i j}$ are distinctively large and the afferent neurons $j$ are spatial neighbors [Fig.~\ref{fig:winput}]. Both the sparsity property and the locality property may be a consequence of the fact that the correlations in the MNIST system are mostly contributed by spatially neighboring pixels. For the MNIST dataset, all the elements $A_{i j}$ of the correlation matrix are non-negative, and non-surprisingly, all the large-magnitude synaptic weights $w_{i j}$ are positive.

Similarly, the lateral projection field of a neuron $i$ is defined as the subset of other neurons $j$ to which the synaptic weights $w_{j i}$ are significantly distinct from zero. Same as the receptive fields, the projection field of a neuron is also sparse and spatially localized [Fig.~\ref{fig:woutput}]. Because of the nonsymmetric property, however, the projection field of a neuron $i$ are not identical to its receptive field. For example, the central pixel $i$ at $(14,14)$ is mostly affected by the pixel $j$ at $(13, 14)$ and the synaptic weight is $w_{i j} = 0.520$, but pixel $i$ affects mostly the two neurons $k$ at $(15, 14)$ and $l$ at $(14, 15)$, with synaptic weights $w_{k i} = 0.570$ and $w_{l i} = 0.571$. Figure~\ref{fig:weightio} also indicates that at the central pixel $i$ there is a strong directional flow of influence from the north side to the eastern and southern sides. Qualitatively similar directional motifs might be common in the biological brain.

\begin{figure*}
\centering
\subfigure[]{
  \includegraphics[width=0.31\linewidth]{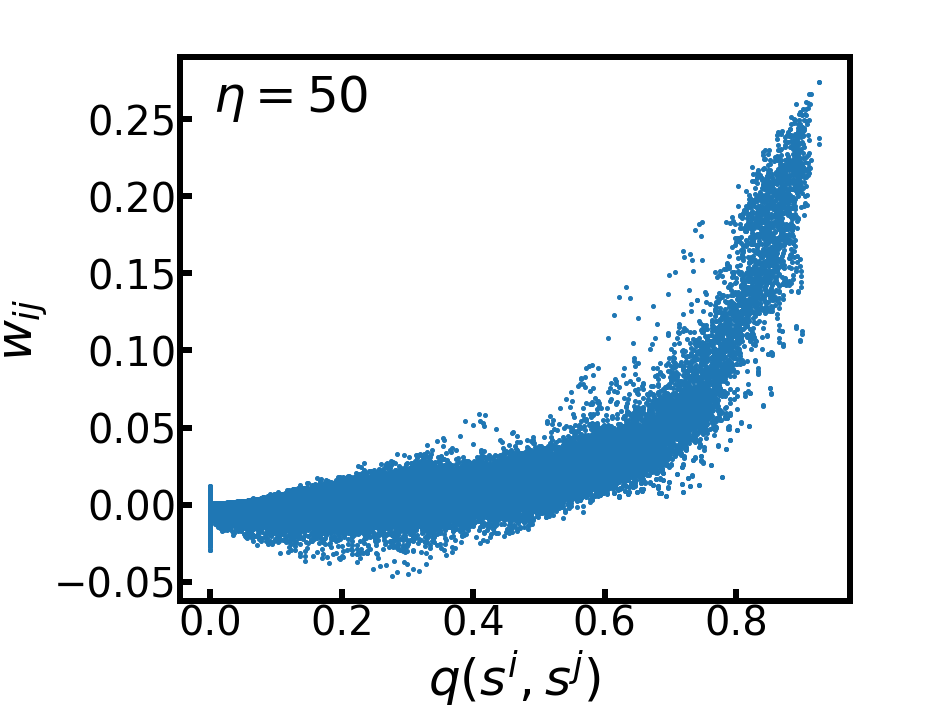}
  \label{fig:wijsij:50}
}
\subfigure[]{
  \includegraphics[width=0.31\linewidth]{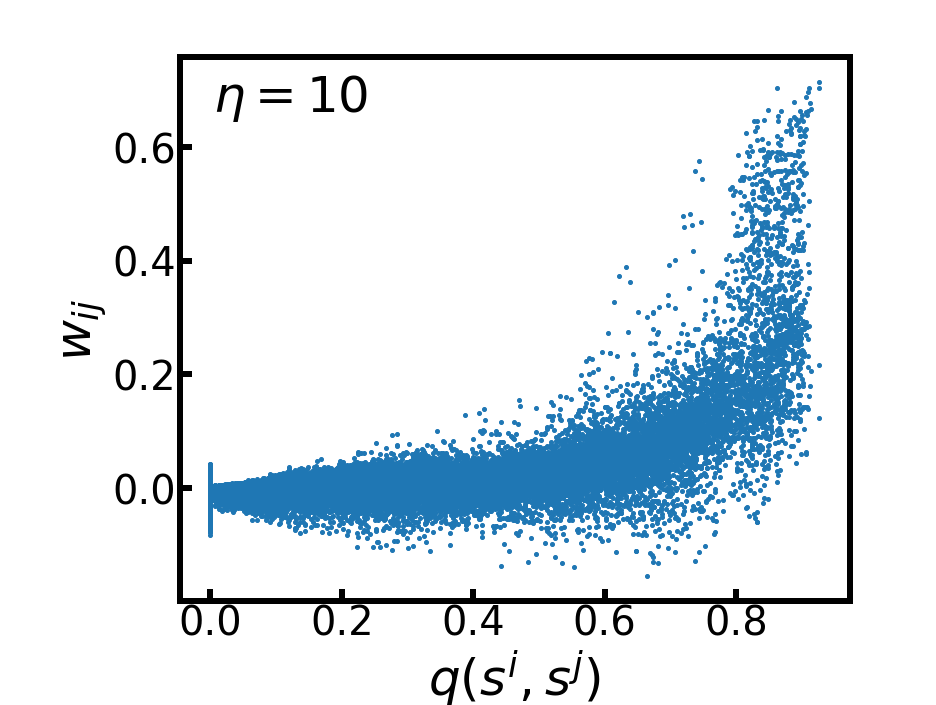}
  \label{fig:wijsij:10}
}
\subfigure[]{
  \includegraphics[width=0.31\linewidth]{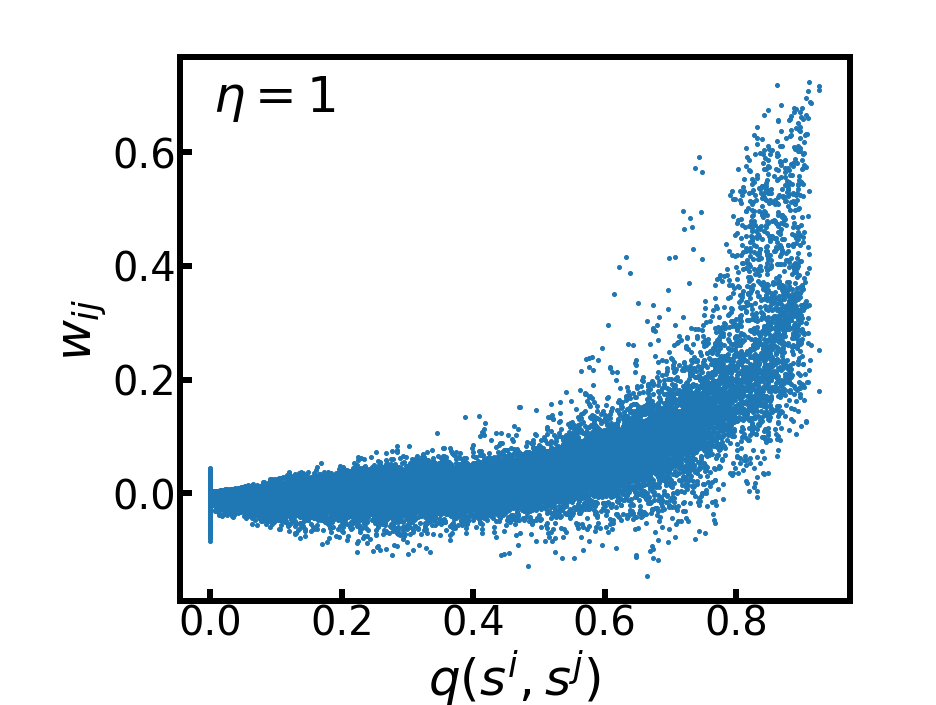}
  \label{fig:wijsij:1}
}
\subfigure[]{
  \includegraphics[width=0.31\linewidth]{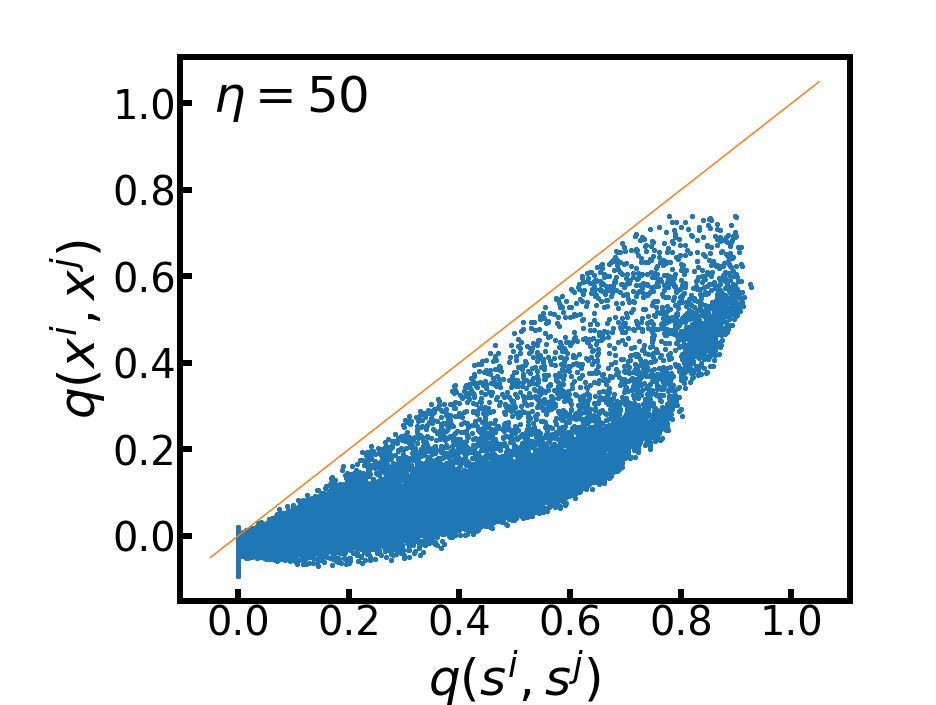}
  \label{fig:xijsij:50}
}
\subfigure[]{
  \includegraphics[width=0.31\linewidth]{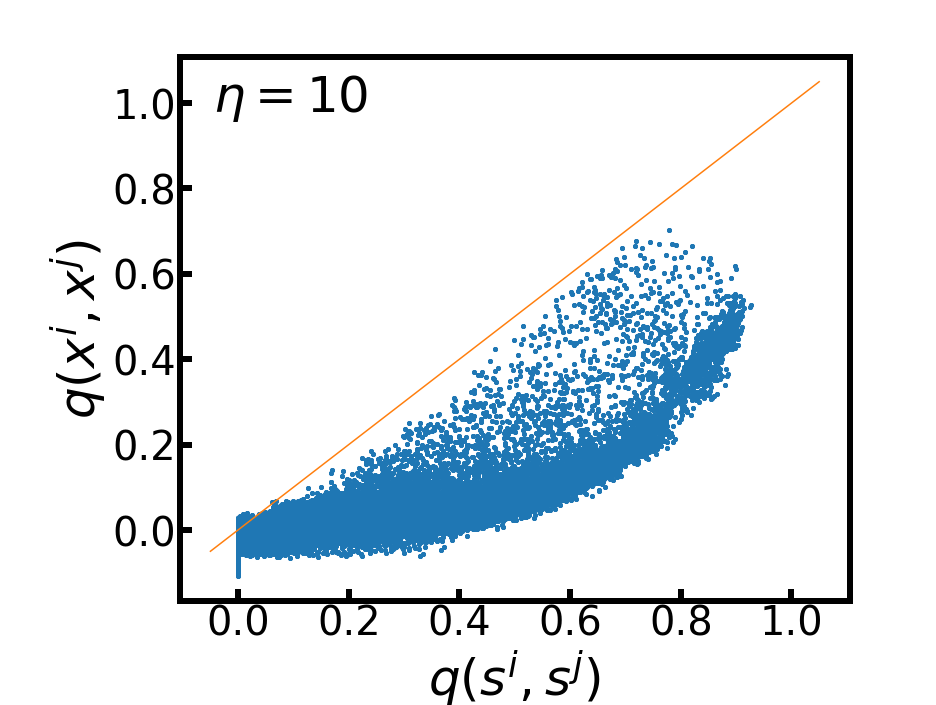}
  \label{fig:xijsij:10}
}
\subfigure[]{
  \includegraphics[width=0.31\linewidth]{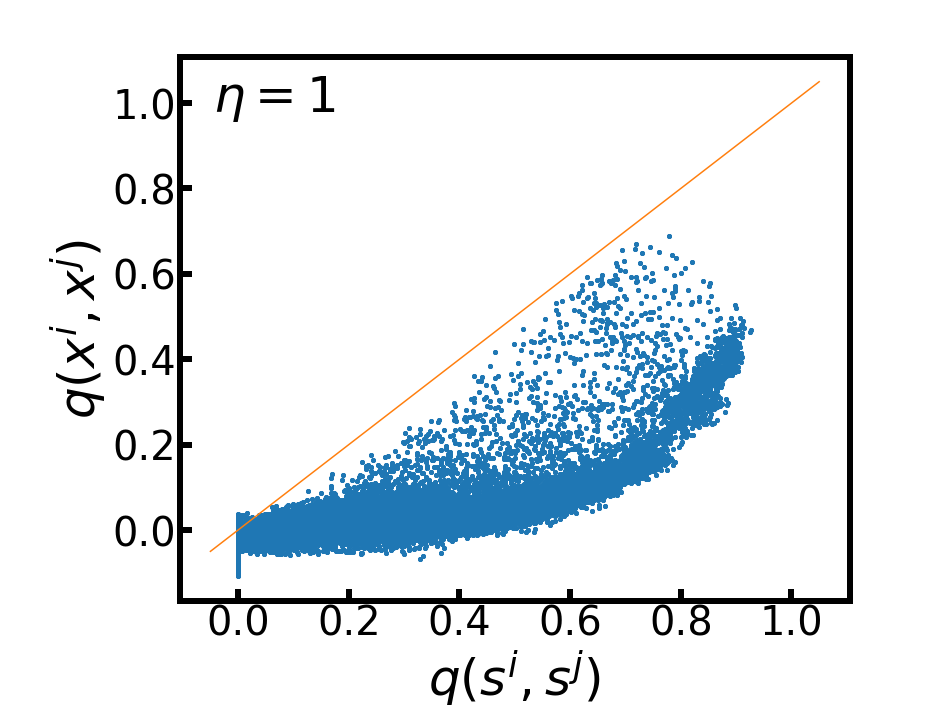}
  \label{fig:xijsij:1}
}
\caption{
Synaptic weights, input correlations and internal correlations. (a-c, top row) Relationship between synaptic $w_{i j}$  and $q(\bm{s}^i, \bm{s}^j)$ (the similarity of input signals at neurons $i$ and $j$). (d-f, bottom row) The relationship between $q(\bm{x}^i, \bm{x}^j)$ (the similarity of internal representations at $i$ and $j$) and input similarity $q(\bm{s}^i, \bm{s}^j)$. The dotted lines mark the hypothetical linear relation $q(\bm{x}^i, \bm{x}^j) = q(\bm{s}^i, \bm{s}^j)$.  In drawing this figure, we only include neurons whose input signals are sufficiently active, that is, the input signals $s_\alpha^i$ for such a neuron $i$ should be nonzero in least $100$ of the $60000$ sample digital patterns $\alpha$. 
}
\label{fig:similarity}
\end{figure*}

For some of the neurons located close to the boundary of the square region, we find that their receptive and projection fields are both empty. In other words, the synaptic weights from other neurons $j$ and to other neurons are both vanishing ($w_{i j} \approx 0$ and $w_{j i} \approx 0$). These neurons are therefore isolated from the other neurons. We find that this isolation is due to the fact that such a neuron $i$ is almost always quiescent ($s_i = 0$).

In natural environment, the visual signals to the retina of an animal are strongly redundant and locally correlated. The locality of the synaptic weights in visual signal processing systems has been well documented~\cite{Srinivasan-etal-1982,Rao-Ballard-1999}. Quantitative experimental investigations on the degree of nonsymmetry $\kappa$ in these lateral interactions may be an interesting experimental issue. It may also be possible that different regions of the cerebral cortex will have different values of the nonsymmetry index $\kappa$. 

\subsection{Neuron pair-wise similarity}

For two $P$-dimensional vector $\bm{s}^i = (s_1^1, \ldots, s_P^1)^\top$ and $\bm{s}^j = (s_1^2, \ldots, s_P^2)$, one on neuron $i$ and the other on neuron $j$, the similarity $q(\bm{s}^i, \bm{s}^j)$ between them is computed as
\begin{equation}
\label{eq:qsisj}
q( \bm{s}^i, \bm{s}^j ) \equiv  \sum\limits_{\alpha = 1}^P 
    \frac{ s_\alpha^i  s_\alpha^j }{ \| \bm{s}^i \| \ \| \bm{s}^j \|} \; .
\end{equation}
This similarity is related to the input correlation $A_{i j}$ by
\begin{equation}
q(\bm{s}^i, \bm{s}^j) = \frac{P}{ \| \bm{s}^i \| \  \| \bm{s}^j \| } A_{i j} \; .
\end{equation}
In other words, $q(\bm{s}^i, \bm{s}^j)$ is a re-scaled correlation of the input signals at neurons $i$ and $j$. 

The top row of Fig.~\ref{fig:similarity} shows the relationship between the input similarity $q(\bm{s}^i, \bm{s}^j)$ of two neurons $i$ and $j$ and the synaptic weights $w_{i j}$ and $w_{j i}$. There is a clear trend of $w_{i j}$ increasing with $q(\bm{s}^i, \bm{s}^j)$, which is naturally anticipated. A large value of synaptic weight $w_{i j}$ implies a large value of similarity  $q(\bm{s}^i, \bm{s}^j)$. Very interestingly, however, the reverse is not necessarily true. The synaptic weight $w_{i j}$ or $w_{j i}$ (or both) could be very close to zero even if the similarity  $q(\bm{s}^i, \bm{s}^j)$ is quite large. For example, the input similarity of pixel $i$ at $(23,14)$ and pixel $j$ at $(26,11)$ is $q(\bm{s}^i, \bm{s}^j) = 0.238$ while both $w_{i j}$ and $w_{j i}$ are very small ($\approx 2\times 10^{-5}$). This means that the lateral neural network may choose to predict the input signal of a neuron $i$ based on the internal states of a few (but not all) of the most highly correlated neurons $j$. The underlying reason might be the redundancy of information in the input signals. If the input signals of both neurons $j$ and $k$ are good predictors of that of neuron $i$, one of the synaptic weights $w_{i j}$ and $w_{i k}$ may be spared to reduce synaptic energy.

\begin{figure*}
  \centering
  \includegraphics[width=0.7\linewidth]{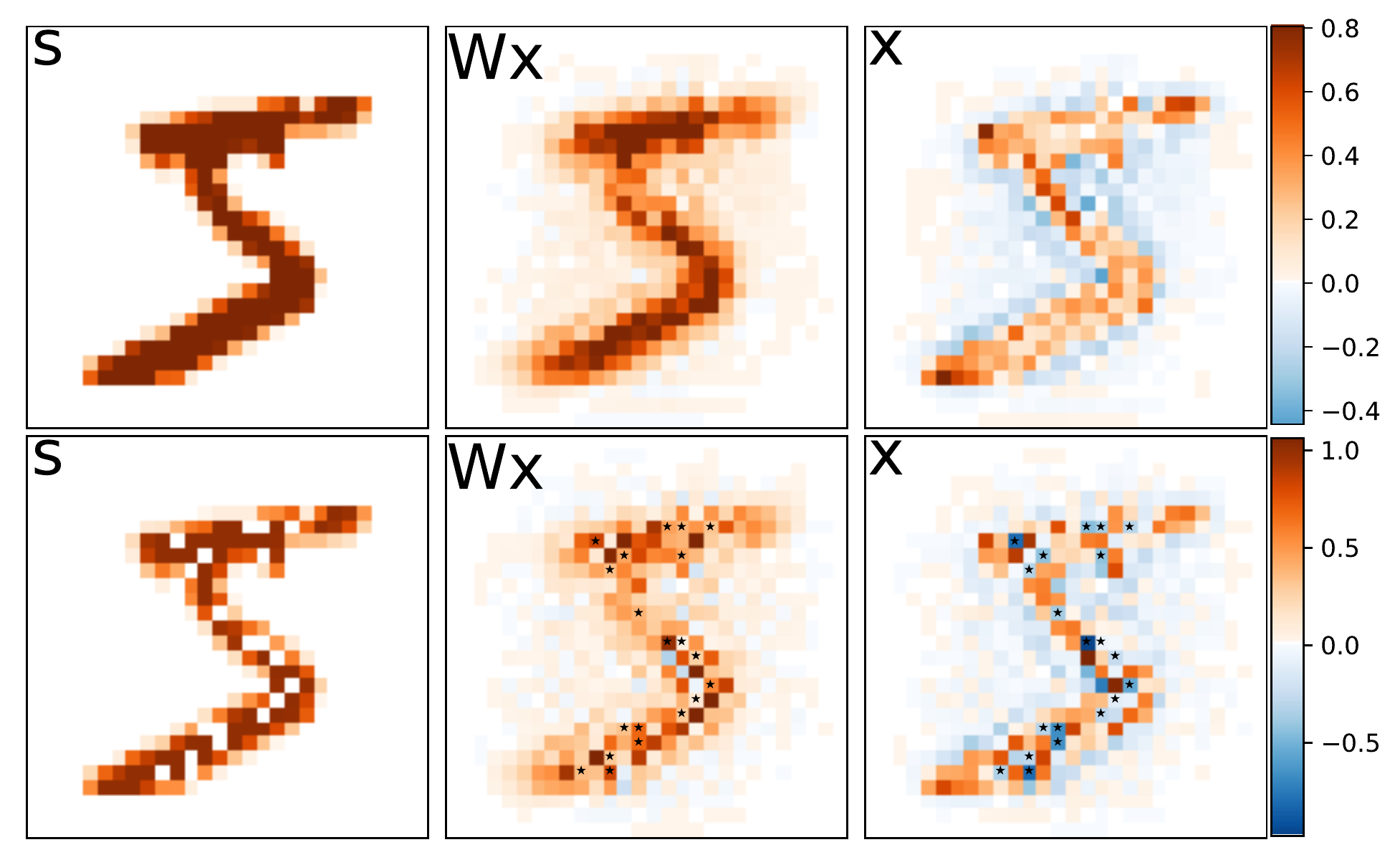}
\caption{
Examples of input signal vectors $\bm{s}_\alpha$ (left column), the predicted signals $\bm{p}_\alpha = \bm{W} \bm{x}_\alpha$ (middle column), and the prediction errors $\bm{x}_\alpha$ (right column). The top left sample is an intact symbol `$5$', while the bottom left sample is an occluded version with some pixels of high intensities being changed to zero intensity (white). The occluded pixels are indicated by a small `$\star$' in the bottom row. The network is trained with weight penalty $\eta = 1$. 
}
\label{fig:digit5}
\end{figure*}

In some sense, the lateral neural network attempts to explain the complicated correlations of the input signal vectors by a few ``direct" interactions. This is similar to recent work on direct coupling analysis in neural sequences and protein sequences, which also tried to distinguish between direct interactions and indirect transmission of correlations~\cite{Schneidman-etal-2006,Cocco-etal-2017}.

The internal states $x_\alpha^i$ of a neuron $i$ depend on the input pattern and they also form a $P$-dimensional vector $\bm{x}^i = (x_1^i, \ldots, x_P^i)^\top$. The similarity $q(\bm{x}^i, \bm{x}^j)$ between the internal vectors of two neurons is
\begin{equation}
    \label{eq:qxixj}
    q(\bm{x}^i, \bm{x}^j) \equiv \sum\limits_{\alpha = 1}^P 
    \frac{ x_\alpha^i  x_\alpha^j }{ \| \bm{x}^i \| \  \| \bm{x}^j \|} \; .
\end{equation}
As demonstrated in the bottom row of Fig.~\ref{fig:similarity}, the internal similarity between two neurons $i$ and $j$  is much smaller than the input similarity between them, that is,
\begin{equation}
\label{eq:qxixjqsisj}
    q(\bm{x}^i, \bm{x}^j) \ < \ 
    q(\bm{s}^i, \bm{s}^j) \; .
\end{equation}
To be more quantitative, the mean value of $q(\bm{s}^i, \bm{s}^j)$ averaged over all the neuron pairs is $0.176$, while the mean value of $|q(\bm{x}^i, \bm{x}^j)|$ is only $0.023$ at $\eta = 1$. Clearly, as a consequence of predictive learning, the correlations among the internal states of different neurons are much reduced in comparison with the strong input correlations. This is a known advantage of predictive coding~\cite{Huang-Rao-2011}.

There are still considerable correlations between the internal states of many neurons and the internal similarities $q(\bm{x}^i, \bm{x}^j)$ between these neurons are quite distinct from being zero.  An interesting idea might be to take the internal state vectors $\bm{x}_\alpha$ as input training signals to another laterally connected layer of predictive-coding neurons. This hierarchical sequence may need to be extended to more layers, until the output vectors are formed by mutually independent elements. By this way, hierarchical predictive coding become a renormalization model~\cite{Mehta-Schwab-2014,Lin-Tegmark-2016b,Bradde-Bialek-2017}. May be only a few elements of the final output vector are significantly different from being zero, and they may offer an obvious classification of the initial input digital pictures. This idea needs to be explored in the future.

\subsection{Surprisal, attention, and prediction}

We present in the top row of Fig.~\ref{fig:digit5} the result of the response dynamics obtained for a randomly chosen image sample $\bm{s}_\alpha$ (a digit $5$). The prediction error ($\bm{x}_\alpha)$ and prediction ($\bm{p}_\alpha \equiv \bm{W} \bm{x}_\alpha$) vectors of this example share some common features with the results obtained on the other samples of the MNIST dataset. First, we find that the predictions $\bm{p}_\alpha$ are visually quite similar with the input signal $\bm{s}_\alpha$. For instance, at weight penalty $\eta = 1$ the similarity between these two $N$-dimensional vectors,
\begin{equation}
q(\bm{s}_\alpha,  \bm{p}_\alpha) \equiv \frac{1}{ \| \bm{s}_\alpha \|  \ \| \bm{p}_\alpha\| }
\sum_{i=1}^{N} s_\alpha^i p_\alpha^i
\; ,
\label{eq:similaritydata}
\end{equation}
has a high value of $0.93 \pm 0.03$, averaged over all the $60000$ digital samples. The optimized synaptic weight matrix $\bm{W}$ could explain the input correlations with high precision.

Second, we observe that the magnitude $| x_\alpha^i|$ of the prediction error is often most significant at the boundary pixels of the original digit symbols, and the spacial gradients of  $x_\alpha^i$ at this boundary pixels are comparatively large and the signs of $x_\alpha^i$ also change at these pixels $i$. In other words, the prediction error vector $\bm{x}_\alpha$ highlights the boundary separating the digital symbol and the background. The neurons correspond to the interior pixels of the symbol and to the regions far-away from the symbol often have lower magnitude of internal responses. The prediction error $\bm{x}_\alpha^i$ is the level of surprise with which a neuron $i$ feels about the input signal $\bm{s}_\alpha$. A relative large magnitude of $x_\alpha^i$ may help the neural system to pay special attention to the pixel corresponding to neuron $i$. This is a simple attention mechanism of novelty detection, and in our present model it does not involve the transfer of top-down messages from the higher hierarchical neural layers.

Prediction and novelty detection are most clearly manifested for input digital signals $\bm{s}_\alpha$ of which some pixels are occluded (namely, the original nonzero input pixel values $s_\alpha^i$ are artificially set to be zero). We find that even with many pixels being occluded, the  network could still offer a highly satisfying prediction ($\bm{p}_\alpha$) for the original intact image (bottom row of Fig.~\ref{fig:digit5}). On the other hand, the prediction errors are the most significant for the occluded pixels, which could guide attention to these regions. These simulation results on occluded input signals  demonstrate clearly the dual role of $\bm{x}_\alpha$ being both an internal model (by combining $\bm{W}$) and being a prediction error vector.

How will a biological brain possibly take advantage of the decomposition (\ref{eq:sdecomp}) to facilitate perception and action? Maybe the prediction $\bm{p}_\alpha$ and the prediction error $\bm{x}_\alpha$ will be transmitted through different paths to different higher-level processing units. The possible biological significance of this needs to be explored more deeply. We notice that there are actually at least two pathways of visual information process in the human brain~\cite{Goodale-Milner-2018}. One of these pathways (the dorsal visual pathway) is responsible for unconscious blindsight~\cite{Weiskrantz-2009}, and a ventral visual pathway is more closely related to conscious perception.

\subsection{Response time variations}

\begin{figure}
  \centering
  \subfigure[]{
    \includegraphics[width=0.465\linewidth]{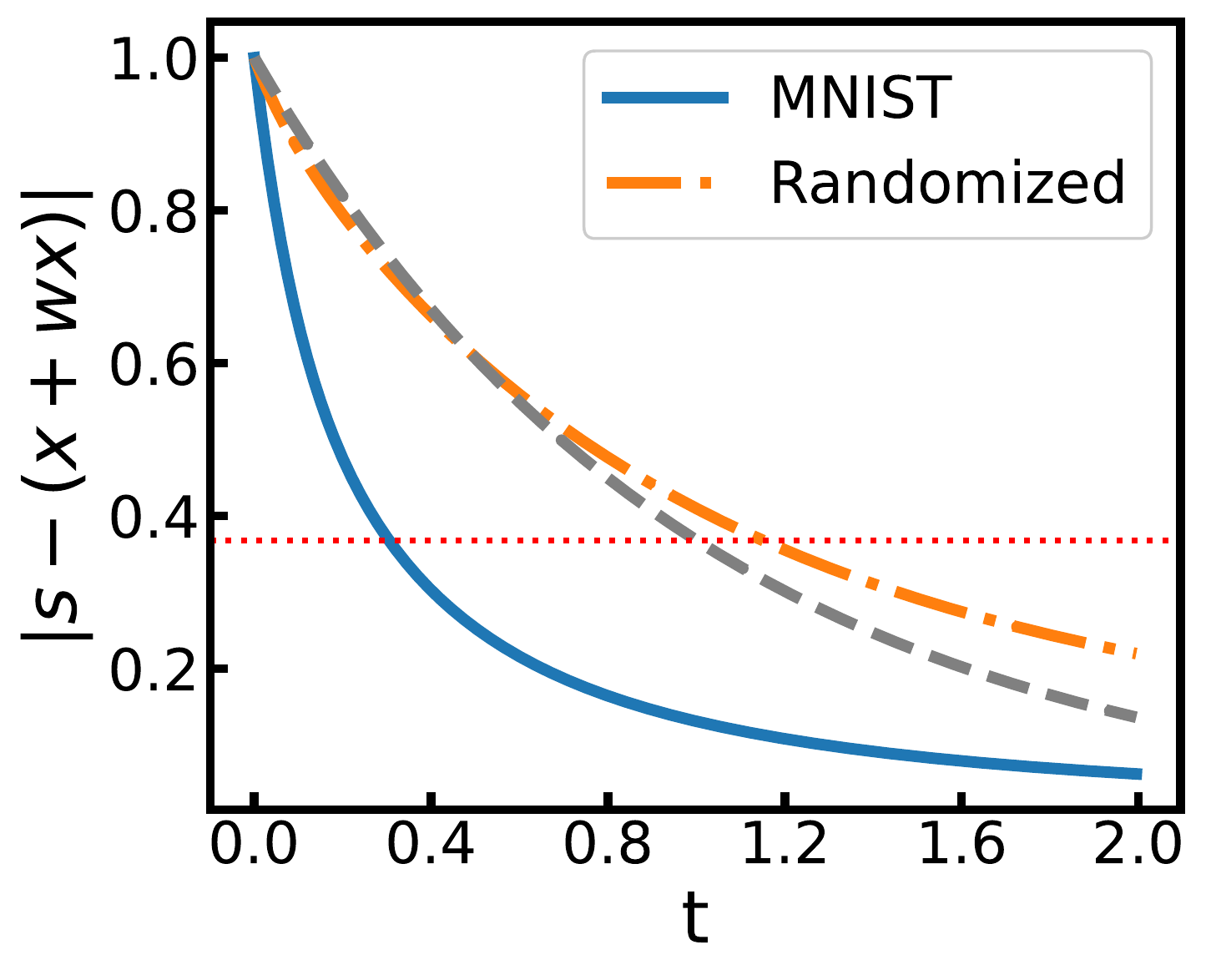}
    \label{fig:restime:A}
  }
  \subfigure[]{
    \includegraphics[width=0.465\linewidth]{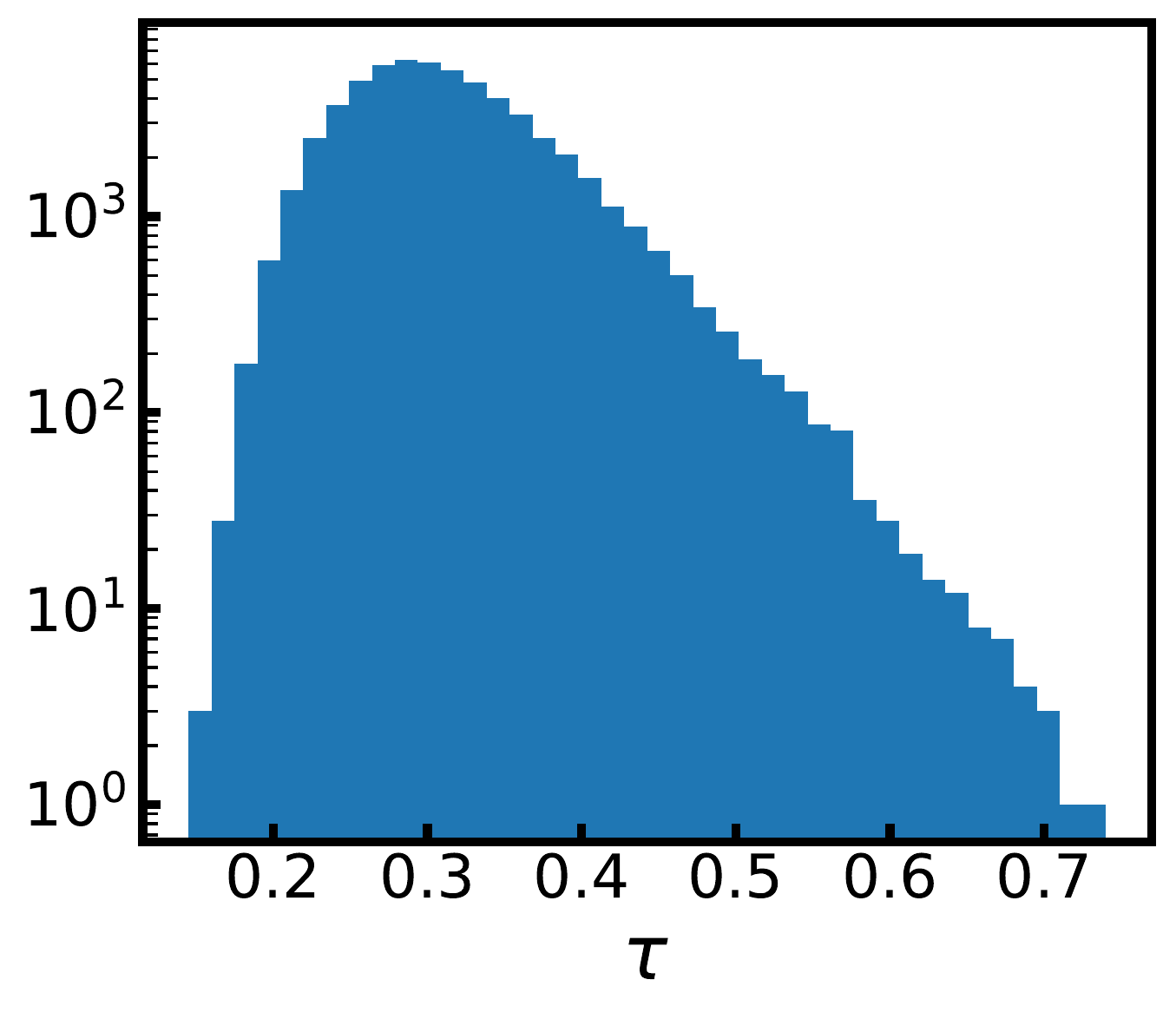}
    \label{fig:restime:B}
  }
  \caption{
    Response time of predictive coding. The weight matrix is trained at penalty value $\eta=1$. (a) The average decay behavior of the magnitude of the difference vector [$\bm{s}-(\bm{x}+\bm{W}\bm{x})$] (rescaled by the magnitude of the input signal $\bm{s}$) wth time $t$. The solid line is obtained for the MNIST images and the dot-dashed line is obtained for the shuffled MNIST images, while the dashed line is obtained for $\bm{W} = \bm{0}$ (no interactions). The horizontal thin line marks the level $1/\textrm{e}$.  (b) The histogram of response times $\tau$ obtained for the MNIST images. 
} 
\label{fig:restime}
\end{figure}

The predictive coding and perception system counteracts an input signal vector $\bm{s}$ by the combined effect of internal state $\bm{x}$ and prediction $\bm{W} \bm{x}$. Figure~\ref{fig:restime:A} reveals the averaged decay behavior of the magnitude of the difference vector [$\bm{s}-(\bm{x} + \bm{W}  \bm{x})$] with time. We define the response time $\tau(\bm{s})$ of the dynamics (\ref{eq:xevol}) to input signal $\bm{s}$ as the earliest time at which the magnitude of the difference vector becomes less than $1/\textrm{e}$ of the initial magnitude $\| \bm{s} \|$. According to Eq.~(\ref{eq:dxdt}), then $\tau(\bm{s})$ is determined by the equation
\begin{equation}
    \frac{ \| \textrm{e}^{-(\bm{I}+\bm{W}) \tau}  \bm{s} \bigl\|  }{ \| 
    \bm{s} \|} = \frac{1}{\textrm{e} } \; .
\end{equation}
If there is no feedback interactions, the response dynamics will be purely exponential and the response time would be the same for any input vector $\bm{s}$, and $\tau(\bm{s})=1$. The response time is much reduced by the introduction of optimized feedback interactions.  For the synaptic weight matrix attained with high penalty ($\eta=50$) the mean response time is $\tau = 0.41 \pm 0.06$ among all the MNIST image samples. This mean response time is further reduced to $\tau = 0.33 \pm 0.06$ at  moderate penalty $\eta=10$ and to $\tau = 0.31 \pm 0.07$ at low penalty $\eta=1$.

For each MNIST image vector $\bm{s}$ we randomly exchange the positions of its elements ($s^i \leftrightarrow s^j$ for pairs of randomly chosen indices $i$ and $j$) and feed the shuffled vector to the network. Very interestingly, we find the response time of the network to such a maximally randomized input is not reduced but rather is increased beyond unity [Fig.~\ref{fig:restime:A}]. This indicates that the recurrent network has the ability to distinguish familiar inputs on which the weight matrix is trained from unfamiliar or novel inputs.

\begin{figure}
  \centering
  \subfigure[]{
    \includegraphics[width=0.95\linewidth]{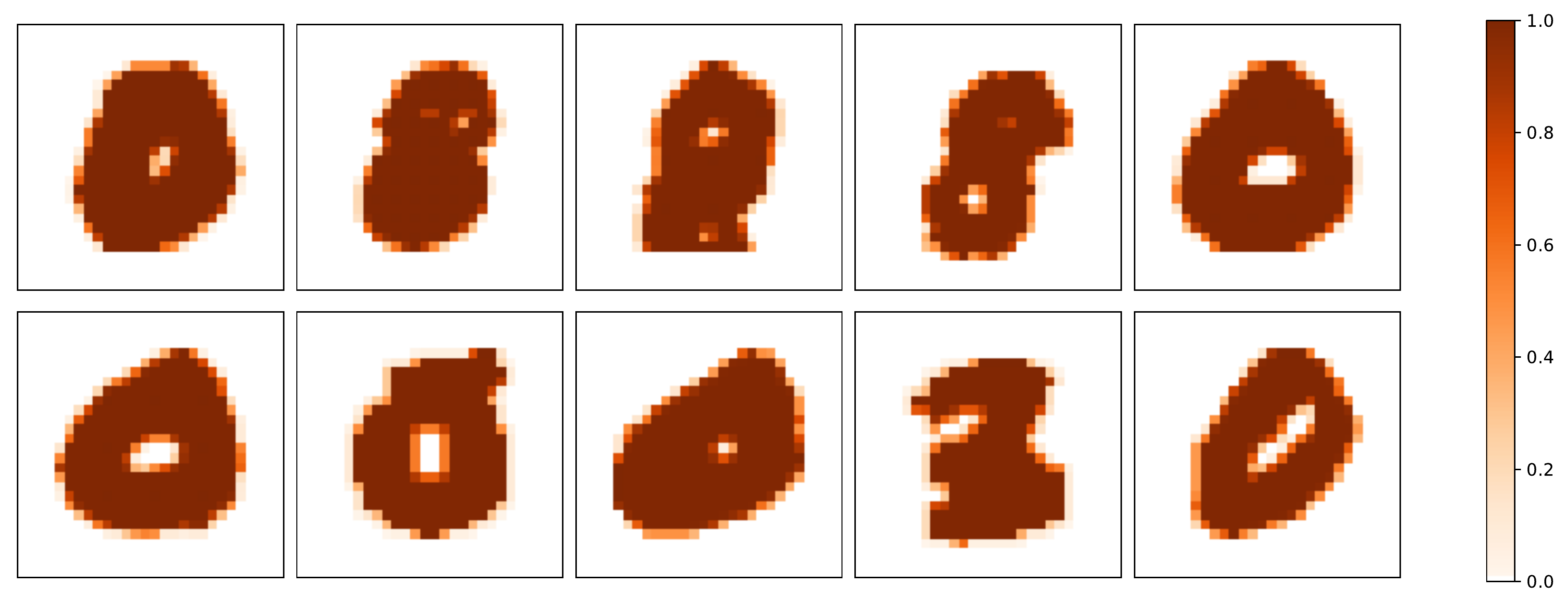}
    \label{fig:extreme:sh}
  }
  \subfigure[]{
    \includegraphics[width=0.95\linewidth]{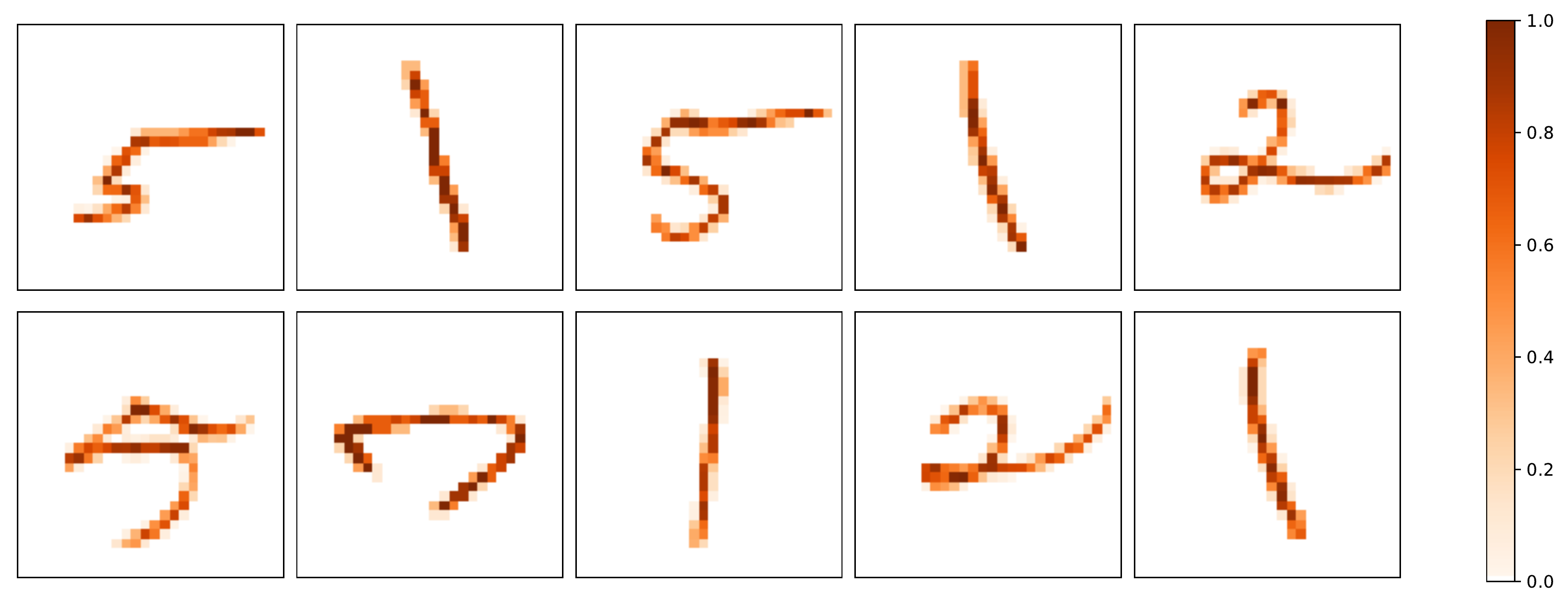}
    \label{fig:extreme:lg}
  }
  \caption{
    Then ten MNIST images with the shortest response times (a), and the ten images with the longest response times (b). The weight matrix is optimized with penalty value $\eta = 1$.
  }
  \label{fig:extreme}
\end{figure}

The response times for the original image vectors also differ considerably, ranging from $\tau = 0.15$ to $\tau =0.73$ at $\eta = 1$ [Fig.~\ref{fig:restime:B}]. The ten images $\bm{s}$ with the shortest response times are shown in Fig.~\ref{fig:extreme:sh}, all of which are found to have very high similarity with the averaged input $\overline{\bm{s}}$, with values $q(\bm{s}_\alpha, \overline{\bm{s}}) > 0.9$. On the other hand, we find that the similarity of the averaged input $\overline{\bm{s}}$ with the leading eigenvector $\hat{e}_1$ of the synaptic weight matrix is very large, $q( \overline{\bm{s}}, \hat{\bm{e}}_1) = 0.9992$ at $\eta = 1$. Then it is easy to understand why these images will be quickly responded by the predictive coding dynamics. The ten images with the longest response times are also shown in Fig.~\ref{fig:extreme:lg}, which are all quite thin and are obviously distinct from the images in Fig.~\ref{fig:extreme:sh}. We find these later images are only weakly aligned with $\hat{\bm{e}}_1$ and $\overline{\bm{s}}$ (the similarity value $q(\bm{s}_\alpha, \overline{\bm{s}}) \approx 0.3$).

The synaptic weights of our network are not trained explicitly to reduce response time. So this elevated response to familiar input signals should be regarded as an extra benefit of predictive perception. The ability to respond quickly to external stimuli is highly desirable in the animal world. The response time $\tau(\bm{s})$ could be used as a measure of typicality of the input vector $\bm{s}$. According to Fig.~\ref{fig:extreme:lg} the input samples with response time $\tau \approx 0.25 - 0.4$ may be regarded as typical inputs, while those with $\tau < 0.2$ or $\tau > 0.6$ may be considered as untypical ones.

\section{Concluding remarks}
\label{sec:conclude}

We studied lateral feedback interactions in a simple model of neural response dynamics (\ref{eq:xevol}) from the perspective of predictive coding. Lateral interactions between two neurons were implemented through the synaptic weights $w_{i j}$ of the linear response function (\ref{eq:linmod}). An optimization problem was formulated to minimize prediction errors, and the method of gradient descent was adopted to evolve the synaptic weights towards near-optimal values. We applied our optimization algorithm to the MNIST dataset of hand-written digits. Our empirical results demonstrated the following four major properties of lateral predictive coding: First, symmetry of interactions is broken in the synaptic weight matrix, with the degree of nonsymmetry $\kappa$ being  significantly positive [Eq.~(\ref{eq:kappa})]; second, the similarity between the internal states $x^i$ and $x^j$ of neurons $i$ and $j$ are significantly reduced as compared to the similarity of the input signals $s^i$ and $s^j$ [Eq.~(\ref{eq:qxixjqsisj})]; third, strong correlations between two neurons $i$ and $j$ do not necessarily mean large synaptic weights between these neurons; and fourth, the response time to familiar input signals is significantly shortened [Fig.~\ref{fig:restime}].

These properties of predictive coding may be highly relevant for information processing in biological neural systems. A natural extension of the present model is a multilayered hierarchical neural network will lateral interactions at individual single layers and feedforward and feedback interactions between adjacent layers. The whole network of the present model could serve as a single layer for a multilayered hierarchical neural information processing system. We did not address the possible effects of lateral interactions in tasks such as data classification and memory retrieval, but these are interesting issues for continued investigations~\cite{Tang-etal-2018,Salvatori-etal-2021,Millidge-etal-2022}.

The linear feedback interactions (\ref{eq:linmod}) is surely too simplistic for biological neurons. The firing rate of a biological neuron is a highly nonlinear and bounded function of the input signals, and the irrelevant information may be lost during the coding and relaying process. Some of the most widely adopted nonlinear functions for theoretical analysis are the logistic function $f_i(x) = 1/(1+\textrm{e}^{-x})$ and the hyperbolic tangent function $f_i(x) = \tanh( x )$~\cite{Pineda-1987,Foldiak-1990}. The introduction of nonlinearity may bring much enhanced competitions among the internal states of different neurons, and consequently it may dramatically affect the learned synaptic weights and change the statistical properties of the internal presentations $\bm{x}$. It may be helpful to start with the extremely nonlinear Heaviside threshold response $f_i(x) = \Theta( x - \theta_i)$, with $\theta_i$ being activation threshold of neuron $i$, to explore the effects of nonlinear lateral interactions in predictive coding.

Another rewarding direction is to consider spiking neurons which are biologically more realistic~\cite{Mikulasch-etal-2022}. The simple noise-free response dynamics (\ref{eq:xevol}) then will be replaced by the more complicated and stochastic  integrate-and-fire dynamics of spiking neurons. Much future work is needed to understand the effect of lateral feedback interactions in predictive coding neural networks and the competition and cooperation between lateral and top-down feedback interactions.

\begin{acknowledgments}

This work was supported by the National Natural Science Foundation of China (Grant Nos.~11975295, and 12047503), and the Chinese Academy of Sciences (Grant Nos.~QYZDJ-SSW-SYS018, and XDPD15). We thank Dr. Qinyi Liao for sharing her GPU computing resource with us.
\end{acknowledgments}


\providecommand{\noopsort}[1]{}\providecommand{\singleletter}[1]{#1}%

\end{document}